\documentclass[aps,preprintnumbers,amsmath,amssymb,prd,superscriptaddress,notitlepage,longbibliography]{revtex4-1}
\pdfoutput=1

\usepackage{graphicx}
\usepackage{dcolumn}
\usepackage{bm}
\usepackage{wasysym}

\newcommand{\frachalf}{\frac{1}{2}}

\newcommand{\kNL}{k_{\rm NL}}

\newcommand{\be}{\begin{equation}}
\newcommand{\ee}{\end{equation}}

\newcommand{\vk}{\mathbf{k}}
\newcommand{\vq}{\mathbf{q}}

\newcommand{\vb}{\mathbf{b}}
\newcommand{\vx}{\mathbf{x}}

\newcommand{\vpi}{\mathbf{\pi}}

\newcommand{\vphi}{\mathbf{\phi}}
\newcommand{\vchi}{\mathbf{\chi}}

\newcommand{\vj}{\mathbf{j}}

\newcommand{\vdelta}{\mathbf{\delta}}

\newcommand{\vp}{\mathbf{p}}
\newcommand{\vep}{\mathbf{\epsilon}}

\newcommand{\vC}{\mathbf{C}}
\newcommand{\vXi}{\mathbf{\Xi}}
\newcommand{\vQ}{\mathbf{Q}}
\newcommand{\vDelta}{\mathbf{\Delta}}

\newcommand{\vN}{\mathbf{N}}
\newcommand{\vA}{\mathbf{A}}
\newcommand{\vM}{\mathbf{M}}

\newcommand{\vF}{\mathbf{F}}

\newcommand{\vpsi}{\mathbf{\psi}}
\newcommand{\vupsilon}{\mathbf{\upsilon}}

\newcommand{\vI}{\mathbf{I}}
\newcommand{\vl}{\mathbf{l}}
\newcommand{\vL}{\mathbf{L}}

\newcommand{\lyaf}{Ly$\alpha$ forest}

\newcommand{\mnras}{{\em Mon. Not. Roy. Astron. Soc. }}
\newcommand{\apjl}{{\em Astrophys. J. Let. }}
\newcommand{\apjs}{ApJS}
\newcommand{\jcap}{JCAP}
\newcommand{\physrep}{{\em Phys. Rept. }}
\newcommand{\aap}{{\em Astron. Astrophys. }}

\newcommand{\pasj}{PASJ}

\newcommand{\anze}{An\v{z}e}

\def\prd{{\em Phys. Rev. }{\bf D }}

\def\pvmhid#1{}

\def\afhid#1{}

\def\ashid#1{}

\newcommand{\dk}{\frac{d^d\vk}{\left(2\pi\right)^d}}
\newcommand{\diffp}{\frac{d^d\vp}{\left(2\pi\right)^d}}
\newcommand{\dq}{d^d\vq}

\begin{document}

\title{Large-scale structure perturbation theory without losing stream 
crossing}

\author{Patrick McDonald}
\email{PVMcDonald@lbl.gov}
\affiliation{Lawrence Berkeley National Laboratory, One Cyclotron Road,
Berkeley, CA 94720, USA}
\author{Zvonimir Vlah}
\email{zvlah@stanford.edu}
\affiliation{Stanford Institute for Theoretical Physics and
Department of Physics, Stanford University, Stanford, CA 94306, USA} 
\affiliation{Kavli Institute for Particle Astrophysics and Cosmology,
SLAC and Stanford University, Menlo Park, CA 94025, USA}

\date{\today}

\begin{abstract}

We suggest an approach to perturbative calculations of large-scale clustering
in the Universe that includes from the start the stream crossing (multiple 
velocities for mass elements at a single position) that is lost in traditional 
calculations.  Starting from a functional integral over displacement, the 
perturbative series expansion is in deviations from (truncated) Zel'dovich 
evolution, with terms that can be computed exactly even for stream-crossed 
displacements. We evaluate the one-loop formulas for displacement and density 
power spectra numerically in 1D, finding dramatic improvement in agreement with
N-body simulations compared to the Zel'dovich power spectrum (which is exact in
1D up to stream crossing).  Beyond 1D, our approach could represent an 
improvement over previous expansions even aside from the inclusion of stream 
crossing, but we have not investigated this numerically.  In the process we 
show how to achieve effective-theory-like regulation of 
small-scale fluctuations without free parameters. 

\end{abstract}

\maketitle

\section{Introduction}

Perturbation theory for large-scale gravitational 
evolution and clustering in the Universe \cite{2002PhR...367....1B} 
should be increasingly valuable as large-scale structure (LSS) surveys become 
increasingly large and precise 
\cite{2014JCAP...05..023F,2016arXiv161100036D}.
From the beginning \cite{1970A&A.....5...84Z}, 
this kind of perturbation theory 
has had a nagging deficiency that there
was no first-principles way to include ``stream crossing''
(often alternatively called ``shell crossing''), i.e., multiple 
different velocities for mass at a single point in space. 
This is a purely technical problem---it is easy
to write down the exact evolution equations for mass elements under effectively 
Newtonian gravity, and easy to see that stream crossing will happen, 
but we simply 
have not had any clean mathematical method to include this phenomenon and still
produce perturbative analytic results for clustering statistics. 
In contrast, a wonderful thing
about N-body simulations \cite{1985ApJ...292..371D,2016MNRAS.463.2273F} 
is they trivially include arbitrarily complex 
velocity structure at a point (pedantically, one can say that you never have
more than one particle at a mathematical point, but we implicitly understand
that the particles in simulations are really an approximation to some 
effectively continuous cloud).  
The key weakness appears in Lagrangian perturbation theory (LPT)
\cite{1993MNRAS.264..375B,1995A&A...296..575B,2008PhRvD..77f3530M,
2011JCAP...08..012O,
2012JCAP...12..004R,2015PhRvD..91b3508V} 
because density  
is approximated by the Taylor expansion of the 
determinant of the local deformation tensor, which is
only correct before stream crossing. 
Eulerian
perturbation theory (EPT) is derived by truncating the evolution 
equations for moments of the velocity distribution function after the 
first moment, i.e., velocity dispersion is set to zero
\cite{1980lssu.book.....P,1986ApJ...311....6G,2011JCAP...04..032M}.  

One unavoidable criticism of various efforts to improve perturbation theory by
effectively summing
to higher orders
\cite{2007PhRvD..75d3514M,2014arXiv1403.7235M,2008ApJ...674..617T}
has been that we were summing a theory that was not
exactly correct anyway, because of missing stream crossing 
\cite{2007PhRvD..75b1302A}.
\cite{2009PhRvD..80d3504P} found only small effects from stream crossing in
numerical simulations; however, this does not mean that missing stream crossing
necessarily has only a small effect on a given perturbation theory calculation. 
One of 
our findings here will be that stream crossing self-regulates, in the following
sense: 
If you ignore it in force calculations, e.g., in the Zel'dovich approximation, 
the extrapolated amount of stream crossing and its effect on the 
density power spectrum is large (the exact Zel'dovich power spectrum does 
include stream crossing for a given 
displacement field). This makes the Zel'dovich power spectrum 
inaccurate, while including the effect of stream crossing on forces, 
feeding back to
suppress itself, greatly improves the results. 
Of course, it is also simply desirable to 
compute small effects to match increasingly high precision observations. 

There have been many efforts to include the effects of stream crossing in
calculations.
References \cite{1977ApJS...34..425D,2011JCAP...04..032M,2011JCAP...10..022T}
included equations for moments of the velocity distribution, but this 
requires a truncation of the moment hierarchy and convergence was never 
demonstrated.  
The adhesion approximation of \cite{1989MNRAS.236..385G,
1988Natur.334..129K,1990MNRAS.247..260W} 
aimed to make crossing streams stop and stick instead of crossing.
Reference \cite{2012JCAP...01..019P} derived evolution equations for explicitly 
smoothed fields in which a coarse-grained velocity dispersion appears, with
contributions to it beyond EPT to be computed by simulations. 
\cite{2012JCAP...07..051B,2012JHEP...09..082C,2014PhRvD..89d3521H} introduced
general ``effective field theory'' counter-terms with free coefficients
that allow fitting for the stream crossing effect in simulations or data.
Reference 
\cite{2016PhRvD..93f3517A} computed the velocity dispersion tensor implied
by LPT. While many of these approaches have had significant success, we would
like to find a more direct calculation. 

Many elements of this paper have appeared before. 
The basic math we exploit to compute forces including stream crossing
has been hiding in plain sight in the calculation of the exact 
Zel'dovich power spectrum (apparently computed first, at least in 
correlation function form, in 
\cite{1988prun.proc..263C}, which we could not find online---see
\cite{2002PhR...367....1B})
which does include stream crossing but only for a
given displacement field.  
Reference \cite{2011A&A...526A..67V} investigated the 
effect of stream crossing by comparing this Zel'dovich calculation to one
where relative streams are stopped ``by hand'' when they would otherwise 
cross. 
Valageas and others have been using 
a similar functional integral formalism for many years
\cite{2001A&A...379....8V,2002A&A...382..412V,
2004A&A...421...23V,2007A&A...465..725V,2007JCAP...06..026M,
2007A&A...476...31V,
2008A&A...484...79V,2012JCAP...03..031S,
2015JCAP...01..014R,2016JCAP...02..032F}. 
The starting formalism of \cite{2016NJPh...18d3020B} is similar 
to ours except for
discussing literal particles instead of a continuum limit, while
\cite{2015PhRvD..91j3507A} is even more similar, although without the 
functional integral formalism. 
Reference \cite{2017MNRAS.470.4858T} has a somewhat similar idea of integrating 
the force after stream crossing. It is generally understood that some form 
of damping of small-scale initial conditions is a good idea 
\cite{2006PhRvD..73f3520C,2006PhRvD..73f3519C,2008PhRvD..77b3533C,
2008PhRvD..78h3503B,2008PhRvD..78j3521B,2011JCAP...06..015A,
2012PhRvD..86j3528T,2012ApJ...760..114S,2012PhRvD..85l3519B}. 
The decisive new feature in this paper is the specific straightforward
perturbative expansion
of the functional integral that we do, allowing concise calculation of 
statistical results. Stream crossing appears essentially effortlessly---in 
fact, if this was the first calculation a person ever saw, they probably 
would not realize stream crossing was a thing to worry about missing at all. 

In the following sections we build up the calculation systematically, 
starting with nothing more than the basic equations for 
Newtonian gravitational evolution. Most of this is basically notation,
that we think 
makes the calculations easier but is not fundamentally connected to the 
inclusion of stream crossing.  
A reader who would like to try to understand the key math trick that we use 
to avoid losing stream crossing 
without learning any
of the formalism may want to read Appendix \ref{app:mathtrick} first, where
we attempt to give a pedagogical taste of what is going on in the main 
calculation.  

\section{Evolution Equations \label{sec:evolutioneq}}

We start with the exact equations for the displacement field $\vpsi(\vq)$ 
of mass elements
labeled by their initial position (Lagrangian coordinate)
$\vq$, at physical position 
$\vx(\vq)=\vq+\vpsi(\vq)$, and the velocity field
$\vupsilon \equiv \vpsi^\prime$, where we use prime to
indicate a derivative with respect to $\eta\equiv \ln a$, which we will
use as our primary
time variable ($a$ is the expansion factor). 
We have, with dots for standard time derivative,
\begin{equation}
\ddot{\vpsi}(\vq)+2 H \dot{\vpsi}(\vq)=\vF\left[\vx\left(\vq\right)\right]
=-\partial_\vx V\left[\vx\left(\vq\right)\right]=
-\partial_\vx V\left[\vq + \vpsi\left(\vq\right)\right]
\label{eq:startingevolution}
\end{equation}
with potential due to density fluctuations
\begin{equation}
V(\vx) = \frac{3}{2} \Omega_m H^2 \partial_\vx^{-2} \delta(\vx) 
\end{equation}
(see \cite{2002PhR...367....1B} for a comprehensive introduction to 
LSS perturbation theory). 
Density at position $\vx$ is
\begin{equation}
1+\delta(\vx) = 
\int \dq ~\delta^D\left[\vx-\vq-\vpsi\left(\vq\right)\right] 
=\int \dq \int \dk 
e^{-i \vk\cdot \left[\vx-\vq-\vpsi\left(\vq\right)\right]}~,
\label{eq:deltax}
\end{equation}
or in (Eulerian) Fourier space
\be
\delta(\vk) = 
\int \dq ~e^{i \vk\cdot \vq} 
\left[e^{i \vk\cdot \vpsi\left(\vq\right)}-1\right] ~.
\label{eq:deltak}
\ee
The density at particle $\vq$, i.e., at position 
$\vx(\vq)=\vq + \vpsi\left(\vq\right)$, is 
\begin{equation}
1+\delta\left[\vx\left(\vq\right)\right] 
=\int \dq^\prime \int \dk 
e^{-i \vk\cdot \left[\vq+\vpsi\left(\vq\right)
-\vq^\prime-\vpsi\left(\vq^\prime\right)\right]}~.
\end{equation}
So finally the force on mass element $\vq$ at $\vx(\vq)$ is
\be
\vF\left[\vx\left(\vq\right)\right]=
-\frac{3}{2}\Omega_m H^2  
\int \dq^\prime \dk \frac{i \vk}{k^2} \left[
e^{-i \vk\cdot \left[\vq+\vpsi\left(\vq\right)-\vq^\prime-
\vpsi\left(\vq^\prime\right)\right]}
-e^{-i \vk\cdot \left[\vq+\vpsi\left(\vq\right)-\vq^\prime\right]}\right] ~,
\label{eq:Fqfull}
\ee
where we have included the second term, coming 
from the mean 
(background) part of the density, subtracted in the definition of
$\delta\equiv \rho/\bar{\rho}-1$, because it definitely does produce a 
force---the force that decelerates the Hubble 
flow---which must be subtracted out
of the first term here. To help clarify: $\vpsi(\vq^\prime)$ here 
represents the displacements
of the mass doing the forcing, while $\vpsi(\vq)$ is the displacement of the
mass being forced, which could just as well be a negligible test mass, which
is why it makes sense for it to appear in the homogeneous part. If
$\vpsi(\vq^\prime)=0$ the forcing field is homogeneous so the peculiar 
acceleration is zero. 

Note that when we write formulas with general dimensionality $d$ here, we
are not really implying a Universe with fundamentally $d$ dimensions. 
We are following
\cite{2016JCAP...01..043M} in modeling a  
3D background Universe with reduced
dimensionality fluctuations, i.e., if $\vpsi(\vq)$ does not depend on a
component of $\vq$, it is easy to see that we can integrate this
component out of all these equations, leaving the reduced $d$ accounting for
only directions that fluctuate. If it is not obvious that the 
qualification that the background Universe is still 3D
matters, note that in a 1D background, with no cosmological constant, in
the Newtonian limit, 
the force between particles is constant, i.e., does not diminish with 
expansion. This means that a truly 1D 
Universe will always turn around and collapse eventually, i.e., there is no
concept of an Einstein-de Sitter (EdS) Universe with power law expansion. 

Specializing to EdS for simplicity, and substituting in $\vupsilon$ to produce 
a 
1st order equation, we can write Eq. (\ref{eq:startingevolution}) as
\begin{equation}
\vupsilon^\prime+\frac{1}{2} \vupsilon - \frac{3}{2}\psi = 
-\frac{3}{2}\partial_\vx \partial_\vx^{-2} \delta 
-\frac{3}{2}\psi
\end{equation}
where note that we have subtracted $\frac{3}{2}\psi$ from both sides to make 
the 
left-hand side correspond to the linearized Zel'dovich evolution, while the
right-hand side is the force beyond this.
We can combine $\vpsi$ and $\vupsilon$ into a single vector 
\be
\vphi\equiv  \left(\begin{array}{c} \vpsi \\ \vupsilon \end{array} \right)
\ee 
and, understanding $\vphi=[\vphi]_i^\alpha(\eta,\vq)$ as a vector
in $\infty\times\infty^d\times d \times 2$
dimensional space, labeled by time, $d$ spatial
coordinates, their standard spatial vector direction,
and $\vpsi$ or $\vupsilon$, write the entire system compactly as
\be
\vL_0 \vphi + \vDelta_0(\vphi) = \vep~, 
\label{eq:compactev}
\ee
where $\vL_0$ is a matrix acting in this 
 $\infty\times\infty^d\times d \times 2$ 
dimensional space, with elements
\be
[\vL_0]_{i_2 i_1}(\vq_2,\eta_2,\vq_1,\eta_1)
= \left( 
\begin{array}{cc}
\frac{\partial }{\partial \eta_2} & - 1\\
-\frac{3}{2} & \frac{\partial}{\partial \eta_2} +\frac{1}{2}\\
\end{array}
\right)\delta^D(\eta_2-\eta_1)\delta^K_{i_2 i_1}
\delta^D(\vq_2-\vq_1)
\ee
[where the explicit 
matrix is over $(\psi,\upsilon)$ and Kronecker-$\delta$ over 
spatial 
directions---note that in Fourier space with coordinate $\vp$ 
this would have $\left(2 \pi\right)^d
\delta^D(\vp_2-\vp_1)$ in place of $\delta^D(\vq_2-\vq_1)$],
\be
\vDelta_0(\vphi) = 
\frac{3}{2}
\left(
\begin{array}{c} 
0 \\
 \partial_\vx \partial_\vx^{-2} \delta + \psi
\end{array}
\right) ~,
\label{eq:Delta0}
\ee
and we have added a stochastic source $\vep=[\vep]_i^\alpha(\eta,\vq)$
with covariance matrix $\left<\vep\vep^t\right>\equiv \vN$
which can be used to set the 
standard initial conditions in the form of an early time impulse. We write
subscript 0 because later the split between $\vL_0$ and $\vDelta_0$ will be 
modified.
$\vep$ in this 
paper will only represent standard differential equation initial conditions, 
so one could ask why bother allowing it to formally have arbitrary time 
dependence. We do this with an eventual Wilsonian renormalization group (RG)
\cite{1974PhR....12...75W}
picture in mind, where small scales are ``integrated out,'' producing       
effectively stochastic differential equations, with $\vep(\eta)$ including 
noise that is not simply initial conditions (in this case we would have
$\vN_0$ and a renormalized $\vN$, as in, e.g., 
\cite{2015JCAP...01..014R,2016JCAP...02..032F}). 
Even for this paper where it is
not strictly needed, we think this formulation is slightly more elegant than
introducing an explicit initial time, when, as we will see, that is 
never necessary.
 
To repeat for clarity: vectors like $\vphi$, $\vDelta$, and $\vep$ are 
generally understood to live in
$\infty\times\infty^d\times d \times 2$ 
dimensional space, labeled by time, $d$ spatial
coordinates, their standard spatial vector direction, 
and $\vpsi$ or $\vupsilon$. 
Matrices like $\vL$ are matrices in this space, and $\vL \vphi$ means a
matrix times vector product, which, if we want to, we can write out in 
explicit coordinates like this:
\be
[\vL \vphi]_{i_2}^{\alpha_2}(\eta_2,\vq_2)
\equiv \int d\eta_1 
\int \dq_1 [\vL]_{i_2 i_1}^{\alpha_2\alpha_1}(\vq_2,\eta_2,\vq_1,\eta_1)
[\vphi]_{i_1}^{\alpha_1}(\eta_1,\vq_1)
\ee 
where $i$ labels vector direction and $\alpha$ labels 
$\vpsi$ or $\vupsilon$...
we generally just write $\vL\vphi$ because it is less tedious. 
Similarly, something like $\vj^t \vphi$ is a dot product in this space, 
producing a scalar. Note that, when it matters (when using Fourier space
coordinates), the $t$ superscript should be
understood to indicate complex conjugation as well as transpose, i.e., 
$\vj^t \vphi \equiv \int \dq~ \vj^t(\vq) \vphi(\vq)\equiv
 \int \diffp \vj^{t}(-\vp) \vphi(\vp)$ (writing out only the 
coordinate part, and 
assuming real fields in $\vq$ space). 
Basically, all of the equations can be understood as standard vector/matrix
equations like you could implement numerically on a computer, if you 
mentally discretize time and space coordinates (integrals and derivatives are
the limits of sums and finite differences as the grid spacing goes to zero). 

Note that
\be
[\vL_0^{-1}]_{i_2 i_1}(\vq_2,\eta_2,\vq_1,\eta_1) = \frac{1}{5}\left[
\left( 
\begin{array}{cc}
3 & 2\\
3 & 2\\
\end{array}
\right) e^{\eta_2-\eta_1}+
\left( 
\begin{array}{cc}
2 & -2\\
-3 & 3\\
\end{array}
\right) e^{-\frac{3}{2}\left(\eta_2-\eta_1\right)} \right]\Theta(\eta_2-\eta_1)
\delta^K_{i_2 i_1} \delta^D(\vq_2-\vq_1)
\ee
(if there is any doubt, multiply this from the left-hand side with $\vL_0$ to 
check). The Heaviside function $\Theta(\eta_2-\eta_1)$ enforces causality of
propagation from  $\eta_1$ to $\eta_2$.

\section{Functional Integral \label{sec:funcint}}

The statistical starting point for our system is that $\vep(\eta,\vq)$ is a 
Gaussian random field with mean zero and correlation $\vN$. We can compute
statistics of interest using a generating function
\begin{equation}
Z(\vj)\equiv \int d\vep~ e^{-\frac{1}{2} \vep^t \vN^{-1} \vep+
\vj^t\vphi[\vep]}~,
\label{eq:epsilonlikelihood}
\end{equation}
i.e., we can pull down powers of $\vphi$ by taking derivatives with respect
to $\vj$ to give averages that we want, e.g., 
\be
\left<\phi\right> \equiv 
\frac{\int d\vep~ \vphi[\vep]  e^{-\frac{1}{2} \vep^t \vN^{-1} \vep}} 
{\int d\vep~ e^{-\frac{1}{2} \vep^t \vN^{-1} \vep}} 
=Z^{-1}(0) 
\left. \frac{\partial Z(\vj)}{\partial \vj}\right|_{\vj=0}~.
\ee
More generally, $N$th order connected correlation functions can be derived by
taking $N$ derivatives of $\ln Z(\vj)$.  
$Z(\vj)$ is simply a mathematical tool encoding all the information necessary 
to compute statistical averages
like this -- by manipulating it we can derive results for all possible averages
at once, rather than computing them piecemeal.
Note that $\vphi[\vep]$ means $\vphi(\eta,\vq)$ depending in principle on 
the full function $\vep(\eta,\vq)$, $\vep$ at all times and positions, 
although causality will of course limit $\vphi$ in practice to depending on
$\vep$ at earlier times. 

Now we change integration variables to $\vphi$ using Eq. (\ref{eq:compactev}), 
to give
\be
Z(\vj)\equiv \int d\vphi~ e^{-S(\vphi)+ \vj^t\vphi}
\ee
with 
\be
S(\vphi) = \frachalf\left[\vL_0 \vphi +\vDelta_0(\vphi)\right]^t \vN^{-1}
\left[\vL_0 \vphi +\vDelta_0(\vphi)\right] ~.
\ee
The Jacobian of the transformation is field-independent so we can drop it
(see Appendix \ref{app:Jacobian}).
$\vN$ is defined such
that $\vL_0^{-1}\vN \vL_0^{-t}=\vC_0$ 
is the standard linear theory power spectrum with its standard time evolution. 

It turns out to simplify calculations to introduce another field which
can be integrated over to produce $S(\phi)$ (up to a field-independent
normalization which is never relevant)
\begin{equation}
\int d\vphi ~e^{-S(\vphi)} \equiv \int d\vphi~ d\vchi ~e^{-S(\vphi,\vchi)}
\label{eq:addchi}
\end{equation}
where
\be
S(\phi,\chi)=
\frachalf \chi^t \vN \chi+i \chi^t \vL_0 \vphi +i \chi^t \vDelta_0(\vphi)~.
\label{eq:Sphichi}
\ee
This just shuffles the nonlinearity into a simpler single term. 
(Recall $\int d\vx \exp\left(-\frachalf\vx^t \vA \vx+\vb^t \vx\right) \propto
\det\vA^{-1/2} \exp\left(\frachalf \vb^t \vA^{-1}\vb\right)$.
We reached this point inspired by the idea of a Hubbard-Stratonovich  
transformation
\cite{1959PhRvL...3...77H,1957SPhD....2..416S}. 
It can also be inspired by the Martin-Siggia-Rose 
formalism \cite{1973PhRvA...8..423M,1977JPhA...10..777P,2004A&A...421...23V}, 
and derived as shown in Appendix 
\ref{app:Jacobian}. )
Note that, ignoring the $\vDelta_0$ term, evaluating the Gaussian integrals 
gives $\left<\phi \phi^t\right>_g=\vC_0$,
$\left<\phi \chi^t\right>_g= -i \vL_0^{-1}$, and 
$\left<\chi \chi^t\right>_g=0$.

\section{Perturbation theory \label{sec:PT}}

The standard perturbative approach to this kind of functional integral
is to split $S$ into a quadratic part that
leads to straightforward Gaussian integrals and a perturbation that we expand
out of the exponential, i.e., with $S=S_g+S_p$, we will do
$\exp(-S)=\exp(-S_g)(1-S_p+S_p^2/2+...)$.
In addition to the obvious move of putting the $\vDelta$ term in $S_p$, we know that small-scale displacements are not well-approximated by Zel'dovich at all
times---generally they are damped. Therefore, it makes no sense to include 
them
in the leading order Gaussian part of the calculation, where they can only 
cause trouble. We can self-consistently suppress them by moving them into the
perturbation term $S_p$, i.e., we define
\be
S_g \equiv 
\frachalf \chi^t \vN \chi+ i \chi^t [W^{-1}\vL_0]\vphi 
\equiv \frachalf \chi^t \vN \chi+i \chi^t \vL \vphi 
\ee
and 
\be
S_p 
\equiv i \chi^t \vDelta_0(\vphi) + i \chi^t [(1-W^{-1}) \vL_0] \vphi 
\equiv i \chi^t \vDelta(\vphi)  
\ee
where $W(k)$ is some simple damping function like 
$W(k) = \exp(-c^2 k^2/2)$.
The key is that now the effective linear propagator 
$\vL^{-1} \equiv \vL_0^{-1} W(k)$ appropriately 
suppresses 
small-scale structure, 
while the term in $S_p$ guarantees that this structure is not
arbitrarily lost---its effects will enter as higher order corrections
through $\vDelta(\vphi) \equiv \vDelta_0(\vphi) + (1-W^{-1}) \vL_0 \vphi$.
We will set $c$ to minimize total higher order correction, i.e., an optimal 
level
of suppression should be the one for which the leading order result is as close
as possible to the final answer.
It is not necessary to think too deeply about this on first reading---we are
just treating small-scale propagation as a perturbation, not fundamentally 
different from how we routinely treat nonlinear interactions as a 
perturbation. 
(We discuss below how we could be much
more sophisticated than this simple Gaussian damping, including modifying
$\vN$, 
but for now we just want to 
be sure to capture the
critical physical effect of generally suppressing high-$k$ fluctuations.)

So we are set up to compute correlations using the generating function:
\be
Z(\vj,\vl)= 
\int d\vphi d\vchi~ e^{-S_g(\vphi,\vchi)+ \vj^t\vphi+\vl^t\vchi} 
(1-S_p + S_p^2/2+...)
\ee
Note that $Z$ is now function of $\vj$ and $\vl$ which allows us to
pull down the $\phi$ or the $\chi$ term.

\subsection{Leading order statistics}

We start at lowest order, keeping only the Gaussian part 
\begin{eqnarray}
Z_0(\vj,\vl)&\equiv&
\int d\vphi~ d\vchi~ e^{-\frachalf \chi^t \vN \chi-i \chi^t \vL \vphi+ 
\vj^t\vphi+\vl^t\vchi} 
\propto 
\int d\vphi~ 
e^{- \frachalf \left(\vphi^t \vL^t+i \vl^t\right) \vN^{-1} 
\left(\vL\vphi+i\vl\right)+ \vj^t\vphi} \\ \nonumber
&=&
\int d\vphi~ 
e^{- \frachalf \vphi^t \vC^{-1} \vphi-
i\vl^t \vN^{-1}\vL \vphi+
\frachalf\vl^t \vN^{-1} \vl+
\vj^t\vphi}
\propto
e^{\frachalf \left(\vj^t-i \vl^t \vN^{-1}\vL \right) \vC
\left( \vj-i \vL^t\vN^{-1} \vl\right)+\frachalf\vl^t \vN^{-1} \vl} \\ \nonumber
&=&
e^{\frachalf \vj^t \vC \vj
-i\vj^t \vL^{-1} \vl}
\end{eqnarray}
where $\vC\equiv \vL^{-1} \vN \vL^{-t}$ is the linearly evolved, damped, 
displacement power spectrum.

Now suppose we want to compute statistics of the density field. 
By construction, we can pull a factor $\vpsi(\vq,\eta)$ out of $Z(\vj)$
using the derivative operator $\partial_{\vj_\vpsi(\vq,\eta)}$ 
and therefore we can pull out 
a factor of $\delta(\vk)$ using the operator
$\int \dq ~e^{i \vk\cdot \vq}
\left[e^{i \vk\cdot \partial_{\vj_\vpsi(\vq)}}-1\right]$.
Generally $\exp(\vx \cdot \partial_\vj)f(\vj) = f(\vj+\vx)$, 
so our 
operator $e^{i \vk\cdot \partial_{\vj_\vpsi(\vq)}}$ adds 
$i \vk \delta^D(\vq-\vq^\prime)$ to $\vj_\vpsi(\vq^\prime)$ within $Z(\vj)$
(time is also an index on all these vectors, 
but we suppress it because it is not doing anything interesting).  

Finally we can compute the density power spectrum: 
\begin{eqnarray}
\left<\delta(\vk_1)\delta(\vk_2)\right>_0 &=& \left.
\int \dq_1 ~e^{i \vk_1\cdot \vq_1}
\left[e^{i \vk_1\cdot \partial_{\vj_\vpsi(\vq_1)}}-1\right]
\int \dq_2 ~e^{i \vk_2\cdot \vq_2}
\left[e^{i \vk_2\cdot \partial_{\vj_\vpsi(\vq_2)}}-1\right]
e^{\frachalf \vj^t \vC \vj} \right|_{\vj=0} \\ \nonumber
&=&
\int \dq_1 ~e^{i \vk_1\cdot \vq_1}
\int \dq_2 ~e^{i \vk_2\cdot \vq_2}
e^{-\frachalf \left[
\vk_1\cdot \vC_\vpsi(0)\cdot \vk_1+ 
2\vk_1\cdot \vC_\vpsi(\vq_1-\vq_2)\cdot \vk_2+ 
\vk_2\cdot \vC_\vpsi(0)\cdot \vk_2\right]
} ~ +k=0~ {\rm piece} \\ \nonumber
&=&
(2\pi)^d \delta^D(\vk_1+\vk_2)
\int \dq ~e^{i \vk_1\cdot \vq}
e^{-\vk_1\cdot\left[
\vC_\vpsi(0)-
\vC_\vpsi(\vq)\right]\cdot \vk_1
} ~ +k=0~ {\rm piece} \\ \nonumber
&=&
(2\pi)^d \delta^D(\vk_1+\vk_2)
\int \dq ~\cos\left(\vk_1\cdot \vq \right)
e^{-\frachalf k_{1 i} k_{1 j} \sigma^2_{i j}(\vq) 
} ~ +k=0~ {\rm piece}
\end{eqnarray}
where $\sigma^2_{i j}(\vq) \equiv 
\left<[\psi_i(\vq^\prime)-\psi_i(\vq^\prime +\vq)]
[\psi_j(\vq^\prime)-\psi_j(\vq^\prime +\vq)]\right>_g$, i.e., the covariance
of relative displacements for points with separation $\vq$, with 
Gaussian weights. This is of course the standard exact Zel'dovich power 
spectrum \cite{1996MNRAS.282..767T}, truncated by $W$. To be clear,
$W$ enters because $\sigma^2_{ij}$ is a Gaussian expectation value, i.e., 
expectation value with weight given by $S_g$, and $W$ is part of $S_g$. 

\subsection{First correction}

Now we consider the first correction.  
\begin{eqnarray}
Z_1(\vj,\vl)&=&
-\int d\vchi~ d\vphi ~
e^{-S_g(\vchi,\vphi)+ \vj^t\vphi+\vl^t\vchi}~ S_p(\vchi,\vphi)
= 
-i\int d\vchi~ d\vphi ~
e^{-S_g(\vchi,\vphi)+ \vj^t\vphi+\vl^t\vchi} 
~\vchi^t \vDelta(\vphi) ~.
\end{eqnarray}
We can do this calculation by manipulating $Z_0(\vj,\vl)$ that we have already
calculated. 

\subsubsection{Nontrivial piece}

The most interesting piece is
\begin{eqnarray}
Z_1^{(a)}(\vj,\vl)&=&
-\frac{3}{2} i \int d\vchi d\vphi 
e^{-S_g(\vchi,\vphi)+ \vj^t\vphi+\vl^t\vchi} 
\vchi^t_\vupsilon \partial_\vx \partial_\vx^{-2} \delta  \\ \nonumber
&=& 
-\frac{3}{2} i \int \dq d\eta \int\dq^\prime \dk \frac{i \vk}{k^2}
\int d\vchi d\vphi 
e^{-S_g(\vchi,\vphi)+ \vj^t\vphi+\vl^t\vchi} 
 \vchi_\vupsilon(\vq,\eta) 
e^{-i \vk\cdot \left[\vq+\vpsi\left(\vq,\eta\right)-\vq^\prime-
\vpsi\left(\vq^\prime,\eta\right)\right]} - {\rm  mean~ part}
\\ \nonumber
&=&
\frac{3}{2}  \int \dq d\eta \int\dq^\prime \dk \frac{\vk}{k^2}
e^{-i \vk\cdot \left(\vq-\vq^\prime\right)}  
\partial_{\vl_\vupsilon(\vq,\eta)}
e^{-i \vk\cdot \left[\partial_{\vj_\vpsi\left(\vq,\eta\right)}-
\partial_{\vj_\vpsi\left(\vq^\prime,\eta\right)}\right]}  
Z_0(\vj,\vl) - {\rm mean~ part}
\\ \nonumber
&=&
-i \frac{3}{2}  \int \dq d\eta \int\dq^\prime \dk \frac{\vk}{k^2}
e^{-i \vk\cdot \left(\vq-\vq^\prime\right)}  
e^{-i \vk\cdot \left[\partial_{\vj_\vpsi\left(\vq,\eta\right)}-
\partial_{\vj_\vpsi\left(\vq^\prime,\eta\right)}\right]}  
\vj^t \vL^{-1}_{*\vupsilon(\vq,\eta)}
Z_0(\vj,\vl) - {\rm mean~ part}
\end{eqnarray}
$\vM_{*\vupsilon}$ means the right index of
$\vM$ is $\vupsilon$ and left is summed over in a product with the
adjacent object as usual.
Note that $\vchi$ was projected to its $\vupsilon$ element when it was 
dotted with $\vDelta_0$ given by Eq. (\ref{eq:Delta0}) (ultimately, because
the force acts to change velocity). 

Note that to compute correlations of observables we do not need $\vchi$ so we
can set $\vl$ to zero after the derivative with respect to it. 
The $Z_2$ calculation 
will just involve inserting another set of these derivatives, and so on.

Now, the action of the derivative operator 
$e^{-i \vk\cdot \left[\partial_{\vj_\vpsi\left(\vq,\eta\right)}
-\partial_{\vj_\vpsi\left(\vq^\prime,\eta\right)}\right]}$
on $f[\vj]$
is to add $-i \vk \left[\delta^D(\vq-\hat{\vq})-
\delta^D(\vq^\prime-\hat{\vq})\right] \delta^D(\eta-\hat{\eta})$
to $\vj_\vpsi(\hat{\vq},\hat{\eta})$. This leads to
\begin{eqnarray}
\label{eq:Z1a}
Z_1^{(a)}(\vj)
=
-i \frac{3}{2}  \int \dq &d\eta& \int\dq^\prime \dk \frac{\vk}{k^2}
e^{-i \vk\cdot \left(\vq-\vq^\prime\right)} \\ \nonumber & & 
\vj^t\vL^{-1}_{*\vupsilon(\vq,\eta)}
e^{-i \vk\cdot\left[\vC_{\vpsi(\vq,\eta)*}
-\vC_{\vpsi(\vq^\prime,\eta)*}\right]
\vj-\frachalf k_i\sigma^2_{ij}(\vq-\vq^\prime,\eta)k_j 
 } Z_0(\vj) \\ \nonumber
&-& {\rm mean~part} 
\end{eqnarray}
where note that
$\vL_{\vpsi\vupsilon}^{-1}(\eta,\eta)=0$ leads to significant simplification.
The ``mean part'' is
\begin{eqnarray}
\label{eq:meanpart}
-i &\frac{3}{2}&  \int \dq d\eta \int\dq^\prime \dk \frac{\vk}{k^2}
e^{-i \vk\cdot \left(\vq-\vq^\prime\right)}  
\vj^t\vL^{-1}_{*\vupsilon(\vq,\eta)}
e^{-i \vk\cdot\vC_{\vpsi(\vq,\eta)*}
\vj-\frachalf k_i\vC^\psi_{ij}(0,\eta)k_j
 } Z_0(\vj)
\\ \nonumber
& & =-d^{-1}\frac{3}{2}
\vj^t \vL^{-1}_{*\upsilon} \vC_{\psi*} \vj~ Z_0(\vj)  ~.
\end{eqnarray}
We will see that for $d=1$ 
this precisely cancels the Zel'dovich force-compensating 
term below, so it will not enter the numerical calculations in this paper, 
although it will need to be included in higher dimensional calculations.  

Note that we generally need the normalized generating function 
$Z(\vj)/Z(0)$. 
If we have perturbatively computed
$Z(\vj)=Z_0(\vj)+Z_1(\vj)+...$ we have perturbatively 
$Z(\vj)/Z(0)=Z_0(\vj) + Z_1(\vj)-Z_1(0) Z_0(\vj)$ where we have used
$Z_0(0)=1$. Of course, $Z_1^{(a)}(0)=0$ here. 

Let us first compute the displacement power spectrum, which will be the
Fourier transform of the displacement correlation function:
\begin{eqnarray}
\left<\psi_{i_1}\left(\vq_1,\eta\right)
\psi_{i_2}\left(\vq_2,\eta\right)\right>^{(a)}_1 &=&
\partial_{j_{\psi_{i_1}\left(\vq_1,\eta\right)}} 
\partial_{j_{\psi_{i_2}\left(\vq_2,\eta\right)}}Z_1^{(a)}\left(\vj\right)
\\ \nonumber
&=&
- \frac{3}{2} \int \dq d\eta^\prime \int\dq^\prime \dk \frac{k_j k_i}{k^2}
e^{-i \vk\cdot \left(\vq-\vq^\prime\right)} 
\left[\vL^{-1}\right]_{\psi\vupsilon}^{i_1 j}
\left(\vq_1-\vq,\eta,\eta^\prime\right)
\\ \nonumber
& &\qquad \qquad \qquad
\left[C_\psi^{i i_2}\left(\vq-\vq_2,\eta^\prime,\eta\right)-
C_\psi^{i i_2}\left(\vq^\prime-\vq_2,\eta^\prime,\eta\right)
\right]
e^{-\frachalf k_m\sigma^2_{mn}(\vq-\vq^\prime,\eta^\prime)k_n 
 }  \\ \nonumber
& & +1\leftrightarrow 2-{\rm mean~part}
\\ \nonumber
&=&
- \frac{3}{2} \int \dq d\eta^\prime \int\dq^\prime \dk 
\cos\left(\vk\cdot \vq^\prime\right)
\left[\vL^{-1}\right]^{i_1 j}_{\psi\vupsilon}
\left(\vq,\eta,\eta^\prime\right)
\frac{k_{j} k_i}{k^2}
\\ \nonumber
& &\qquad \qquad \qquad
\left[C^{i i_2}_\psi\left(\vq_{12}+\vq,\eta^\prime,\eta\right)-
C^{i i_2}_\psi\left(\vq_{12}+\vq+\vq^\prime,\eta^\prime,\eta\right)
\right]
e^{-\frachalf k_m\sigma^2_{mn}(\vq^\prime,\eta^\prime)k_n 
 }  \\ \nonumber
& & +1\leftrightarrow 2-{\rm mean~part}
\label{eq:psicorrelation}
\end{eqnarray}
where in the last step we have shifted some variable definitions around
and defined $\vq_{12}\equiv \vq_1-\vq_2$ (when we use one index label when 
there should be two we mean to duplicate the index value, e.g., 
$C_\psi\equiv C_{\psi\psi}$).  
Note that symmetry guarantees that the result, like $C_\psi^{ij}(\vq_{12})$, 
takes the form
$\delta^K_{i j} f_1(q_{12}) + \hat \vq^i_{12} \hat \vq^j_{12} f_2(q_{12})$, 
e.g., if we align the coordinates along the direction of $\vq_{12}$, the 
matrix
is diagonal with value $f_1+f_2$ along the direction of $\vq_{12}$ and $f_1$
in the transverse directions. 
 
We do not write out the mean part because in the 1D calculations in this 
paper it is exactly canceled by the 
Zel'dovich-compensating piece below. Higher dimension calculations will
need to include it.  

We Fourier transform this with respect to $\vq_{12}$ 
to compute the power spectrum:
\begin{eqnarray}
&P^{(a)i_1 i_2}_{\psi}&(\vk_1,\eta)= \\ \nonumber
& &- \frac{3}{2}~\int d\eta^\prime 
\left[\vL^{-1}\right]^{i_1 j}_{\psi\vupsilon}
\left(\vk_1,\eta,\eta^\prime\right)
P^{g i i_2}_{\psi}\left(\vk_1,\eta^\prime,\eta\right)
\int
\dq^\prime 
\left[1-\cos\left(\vk_1\cdot\vq^\prime\right)\right]
\int \dk
\frac{k_j k_i }{k^{2}}
\cos\left(\vk\cdot \vq^\prime\right)
e^{-\frachalf k_m \sigma^2_{mn}(\vq^\prime,\eta^\prime)k_n}  
\\ \nonumber
& & +1\leftrightarrow 2-{\rm mean~part} ~,
\end{eqnarray}
where $P^{g}(\vk)$ is the Fourier transform of 
$C(\vq_{12})$.

For simplicity of obtaining numerical results in this paper, we will specialize
to 1D where it is easy to do the $\vk$ integral: 
\begin{eqnarray}
P_\psi^{(a)1D}(k,\eta)&=&
- 3 \int d\eta^\prime   
\left[\vL^{-1}\right]_{\psi\vupsilon}
\left(k,\eta,\eta^\prime\right) 
P_\psi^g\left(k,\eta^\prime,\eta\right)
\int dq^\prime
\left[1- \cos\left(k q^\prime\right) \right]
\frac{e^{-\frachalf \frac{ q^{\prime 2}}{\sigma^2(q^\prime,\eta^\prime)}}}
{\sqrt{2 \pi \sigma^2(q^\prime,\eta^\prime)}} 
 -{\rm mean~part}~.
\end{eqnarray}
In the low-$k$ limit this becomes 
\begin{eqnarray}
P_\psi^{(a)1D}(k\rightarrow 0,\eta)=
- \frac{3}{2}k^2 P^g_\psi(k,\eta) \int d\eta^\prime   
\left[\vL_0^{-1}\right]_{\psi\vupsilon}
\left(\eta,\eta^\prime\right) 
e^{\eta^\prime-\eta}
\int dq^\prime
q^{\prime 2}
\frac{e^{-\frachalf \frac{ q^{\prime 2}}{\sigma^2(q^\prime,\eta^\prime)}}}
{\sqrt{2 \pi \sigma^2(q^\prime,\eta^\prime)}} 
 -{\rm mean~part}
\label{eq:lowkdamp}
\end{eqnarray}
i.e., for a given power spectrum the integrals over $q^\prime$ and 
$\eta^\prime$ give some damping scale. 
In addition to making calculations more straightforward, 1D is an especially 
clean test case for the introduction of stream crossing effects because the
Zel'dovich approximation is exact in 1D up to stream crossing
\cite{1973Ap......9..144D,2016JCAP...01..043M}---any deviation at all from
Zel'dovich must be a stream crossing effect.
 
We will do calculations for power law initial conditions 
with $\pi^{-1} k ~P(k)=\left(k/\kNL\right)^{n+1}$, with 
$n=$(2, 1, 0.5, 0), because 
\cite{2016JCAP...01..043M} ran 1D N-body simulations for these slopes. 
For $n=0$, with no high-$k$ suppression ($W(k)=1$) this boils down to
$P_\psi^{(a)1D}(k\rightarrow 0,\eta=0)=
- \left(1.85 \frac{k}{\kNL}\right)^2 P^g_\psi(k,0)$, while for $n=1/2$ it is
$P_\psi^{(a)1D}(k\rightarrow 0,\eta=0)=
- \left(1.26 \frac{k}{\kNL}\right)^2 P^g_\psi(k,0)$, i.e., intuitively 
reasonably, the damping scale roughly corresponds to the nonlinear scale. 
For $n\geq 1$ $\sigma^2(q)$ diverges with no small-scale suppression 
so we will discuss those
below, after computing the small-scale restoring term.  

To be clear: there is no correction like this in LPT, where the displacement
power simply follows linear theory, i.e., the Zel'dovich approximation. The
effect here comes entirely from stream crossing. The equation makes intuitive 
sense, with 
$e^{-\frachalf \frac{ q^{\prime 2}}{\sigma^2(q^\prime,\eta^\prime)}}$
representing the probability that elements with Lagrangian 
separation $q^\prime$ have crossed at time $\eta^\prime$.  

\subsubsection{Canceling the Zel'dovich force}

We now compute the Zel'dovich force term that we need to subtract because it is 
included in the Gaussian part:
\begin{eqnarray}
Z_1^{(b)}(\vj,\vl)=-\frac{3}{2} i \int d\vchi ~d\vphi ~
e^{-S_g(\vchi,\vphi)+ \vj^t\vphi+\vl^t\vchi} 
~\vchi_\vupsilon^t \vpsi 
&=& 
-\frac{3}{2} i \partial_{\vl_\vupsilon}^t
\partial_{\vj_\vpsi}
Z_0(\vj,\vl) =
-\frac{3}{2} 
\vj^t \vL^{-1}_{*\upsilon} \vC_{\psi*} \vj~ Z_0(\vj)
\end{eqnarray}
where at the end we have set $\vl=0$. 
We see that for $d=1$ this exactly cancels the ``mean part'' of 
$Z_1^{(a)}$ [Eq. (\ref{eq:meanpart}), 
which is subtracted from $Z$, so it  appears with a positive 
sign]. If it exactly cancels in $Z$, it will exactly cancel in all 
statistics. 

For higher dimensions the cancellation against
Eq. (\ref{eq:meanpart}) is only partial, so the 
contribution of this term to statistics must be computed.
It is interesting to 
compute the time dependence of, e.g., the 
$\vpsi(\vq_1,\eta)$--$\vpsi(\vq_2,\eta)$ correlation function
contribution, 
$\propto \int \dq d\eta^\prime 
\vL^{-1}_{\vpsi\vupsilon}(\vq_1,\eta,\vq,\eta^\prime)
\vC_{\vpsi\vpsi}(\vq,\eta^\prime,\vq_2,\eta) \propto
e^{2 \eta}\left(\eta-\eta_i-2/5\right)$ where we have included an initial 
time $\eta_i$ in the term where it does not give zero in the 
$\eta_i\rightarrow -\infty$ (expansion factor $a_i\rightarrow 0$) limit.
This is not a problem because it cancels a similar term coming out of the 
other part of Eq. (\ref{eq:Z1a}), 
to produce a final result that is insensitive to
the initial time. We just need to make sure to do numerical 
calculations in a
way that preserves this cancellation. 

\subsubsection{Restoring the suppressed small-scale fluctuations}

Finally we compute the term restoring the small-scale fluctuations that we
suppressed in the Gaussian part:
\begin{eqnarray}
Z_1^{(c)}(\vj,\vl)= -i \int d\vchi d\vphi 
e^{-S_g(\vchi,\vphi)+ \vj^t\vphi+\vl^t\vchi} 
\vchi^t (1-W^{-1})\vL_0 \vphi 
&=& 
-i \partial_\vl^t(1-W^{-1}) \vL_0 \partial_\vj
Z_0(\vj,\vl) \\ \nonumber
&=& \left[ {\rm constant} + \vj^t \left(1-W\left(k\right)\right)
\vC \vj \right]Z_0(\vj) ~,
\end{eqnarray}
where we again set $\vl=0$ at the end because we do not need $\vl$ to compute
observable statistics. 
This is adding back suppressed power, at lowest order. The constant is an
irrelevant normalization factor. 

The displacement correlation contribution is simply 
\be
\left<\psi_{i_1}\left(\vq_1,\eta\right)
\psi_{i_2}\left(\vq_2,\eta\right)\right>^{(c)}_1 =
2~\vC^{\psi \left(1-W\right)}_{i_1 i_2}\left(\vq_1-\vq_2,\eta\right)
\ee
where $\vC^{\psi (1-W)}$ is understood to mean $\vC^{\psi\psi}$ computed
with a factor $1-W(k)$ multiplying the power spectrum, i.e., the 
corresponding power spectrum contribution is
\be
P^{(c)}_{\vpsi}(\vk,\eta) =
2\left[1-W\left(k\right)\right] P^g_{\vpsi}(\vk,\eta) ~.
\ee

\begin{figure}[ht!]
 \begin{center}
  \includegraphics[scale=0.8]{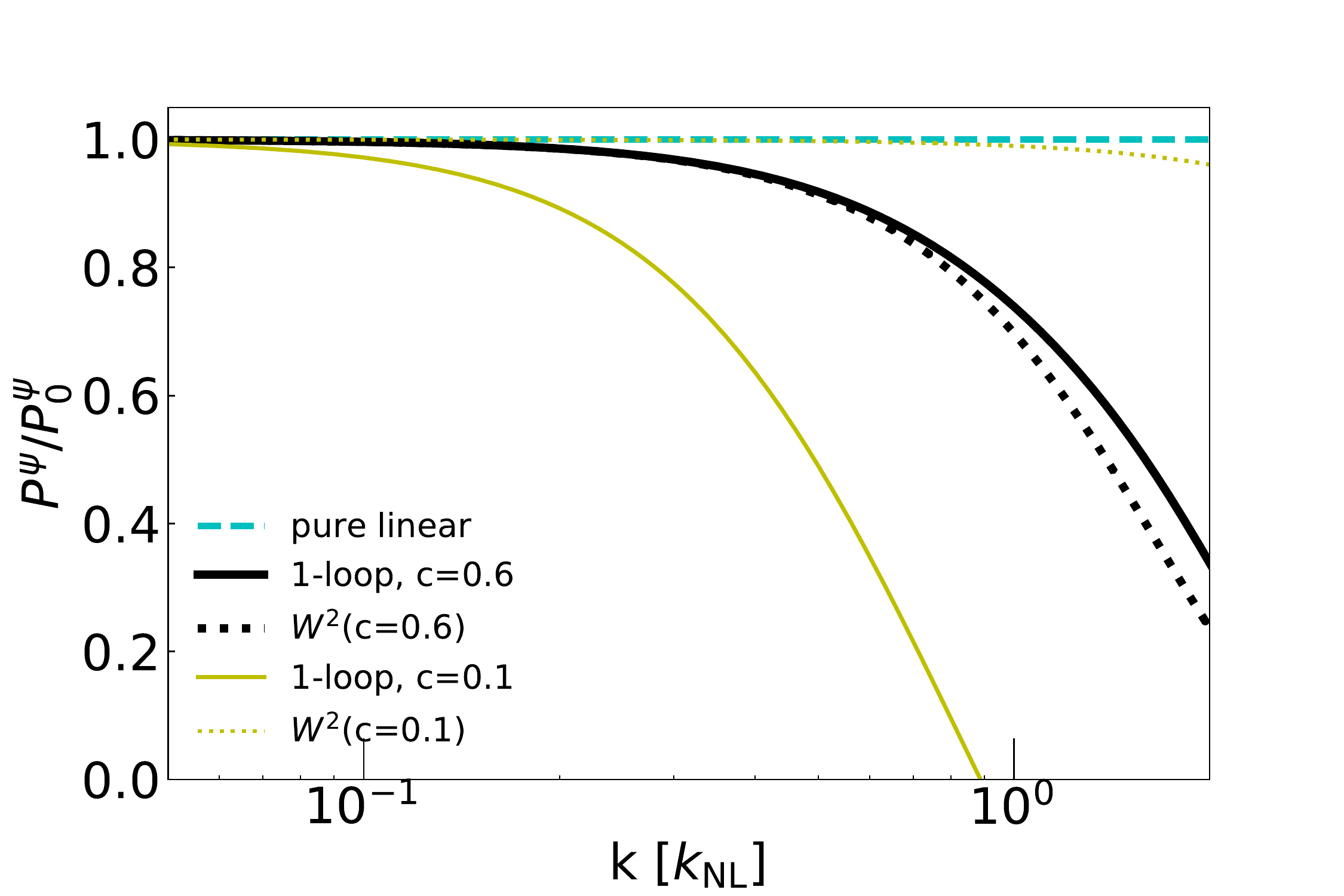}
 \end{center}
 \caption{
One loop suppression of displacement power, in 1D, for $n=2$.
Dotted lines 
show the suppression of the linear power by 
$W(k)\equiv\exp\left(-k^2 c^2/2\right)$, for the value of
$c$ that sets the one loop
correction to zero in the $k\rightarrow 0$ limit (black) or a smaller value
for contrast (yellow), i.e., these are effectively truncated Zel'dovich, 
while the denominator is bare Zel'dovich. 
Solid adds the
1-loop correction, i.e., dotted is the leading order result and the
difference between solid and dotted is the correction.
For $n=2$ the correction in the $c\rightarrow 0$ limit is rapidly divergent. 
Note that the preferred $c$ is
fixed entirely by the principle that the correction goes to zero in the 
low-$k$ limit---it is not a free parameter. In fact, we do not have 
simulations of the 1D displacement power, so the black curve is literally a
(lowest order) prediction. 
}
 \label{fig:Ppsi2}
\end{figure}

\begin{figure}[ht!]
 \begin{center}
  \includegraphics[scale=0.8]{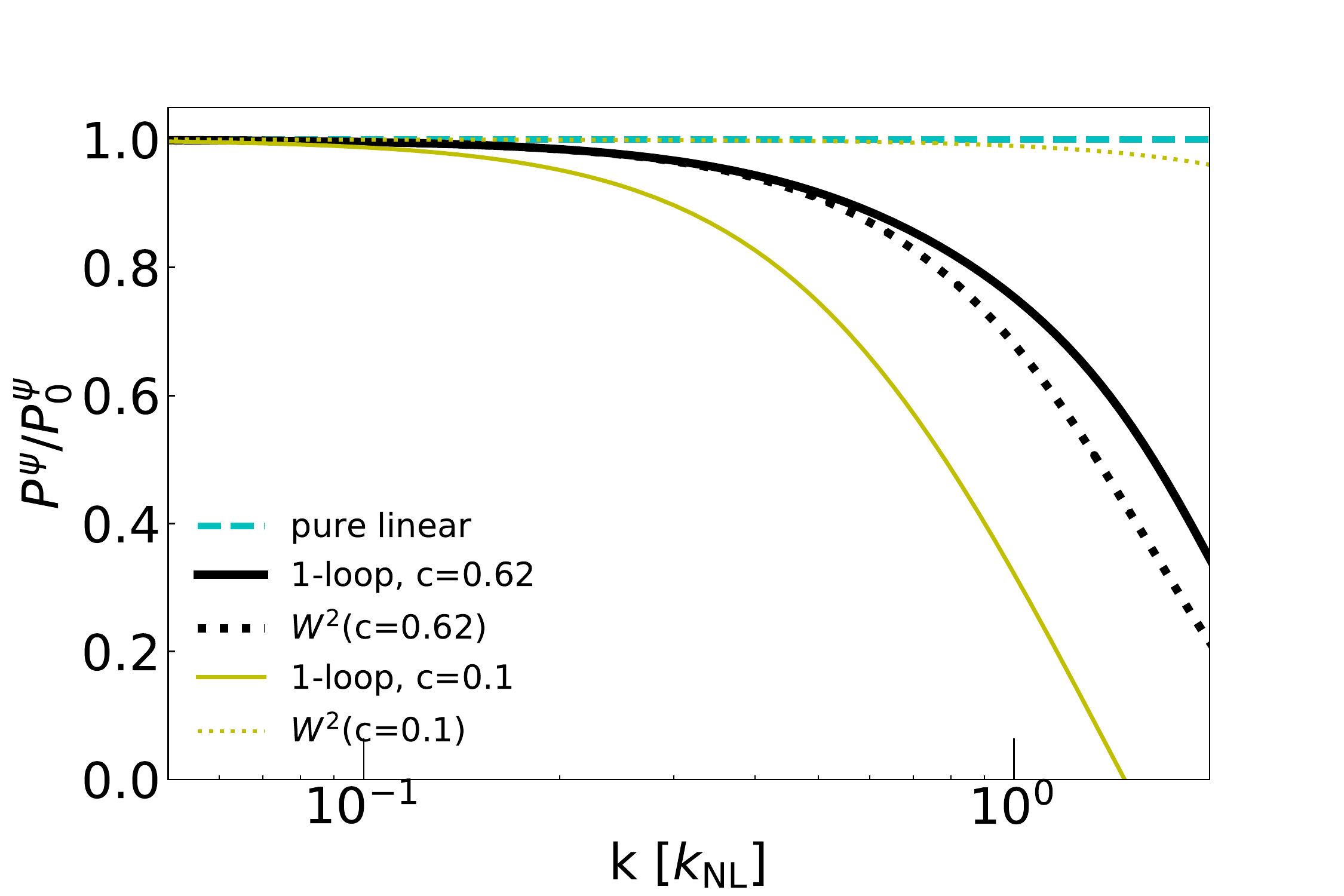}
 \end{center}
 \caption{
One loop suppression of displacement power, in 1D, for $n=1$, 
similar to Fig. \ref{fig:Ppsi2}.  
 }
 \label{fig:Ppsi1}
\end{figure}

\begin{figure}[ht!]
 \begin{center}
  \includegraphics[scale=0.8]{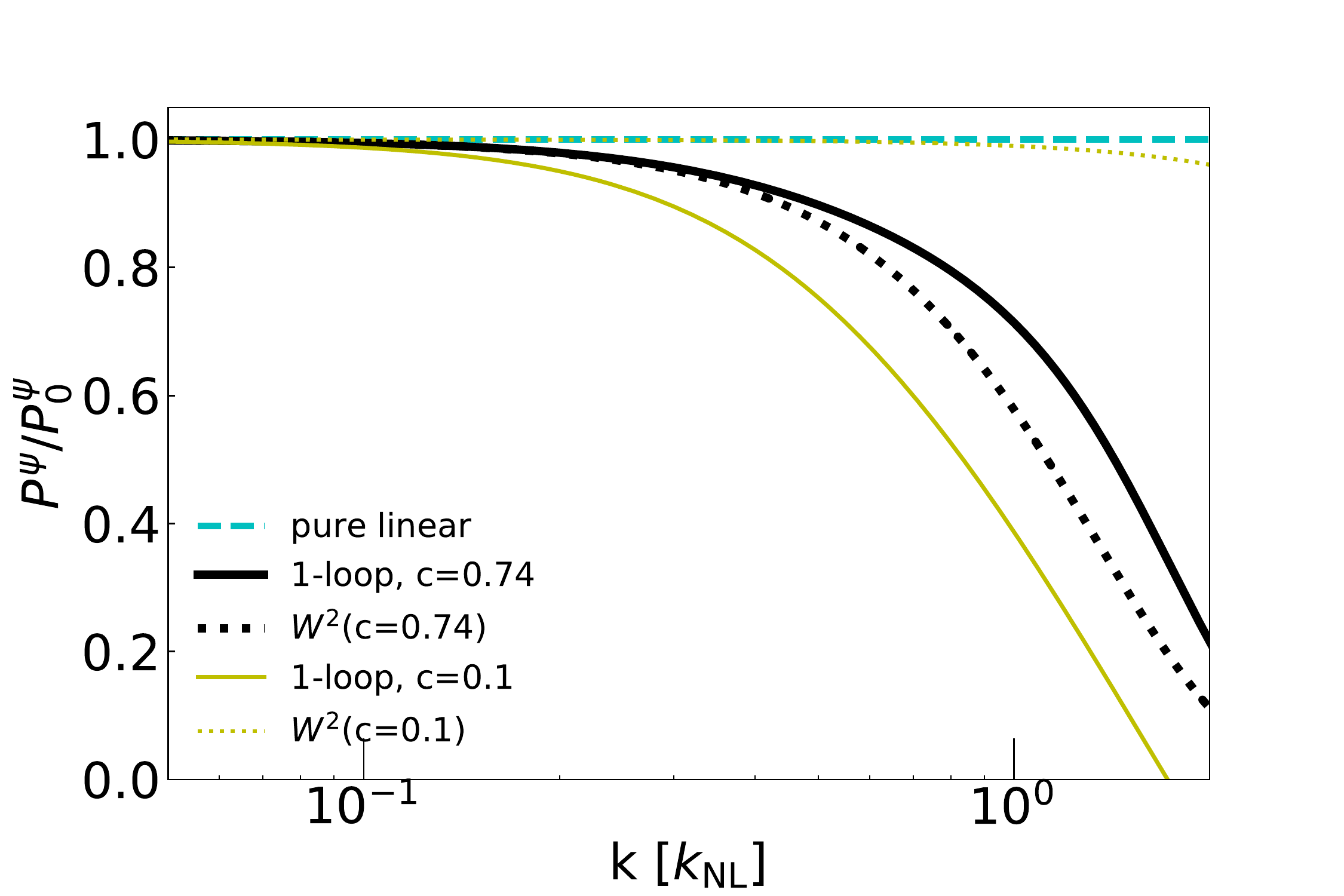}
 \end{center}
 \caption{
One loop suppression of displacement power, in 1D, for $n=0.5$, 
similar to Fig. \ref{fig:Ppsi2}.
 }
 \label{fig:Ppsi0.5}
\end{figure}

\begin{figure}[ht!]
 \begin{center}
  \includegraphics[scale=0.8]{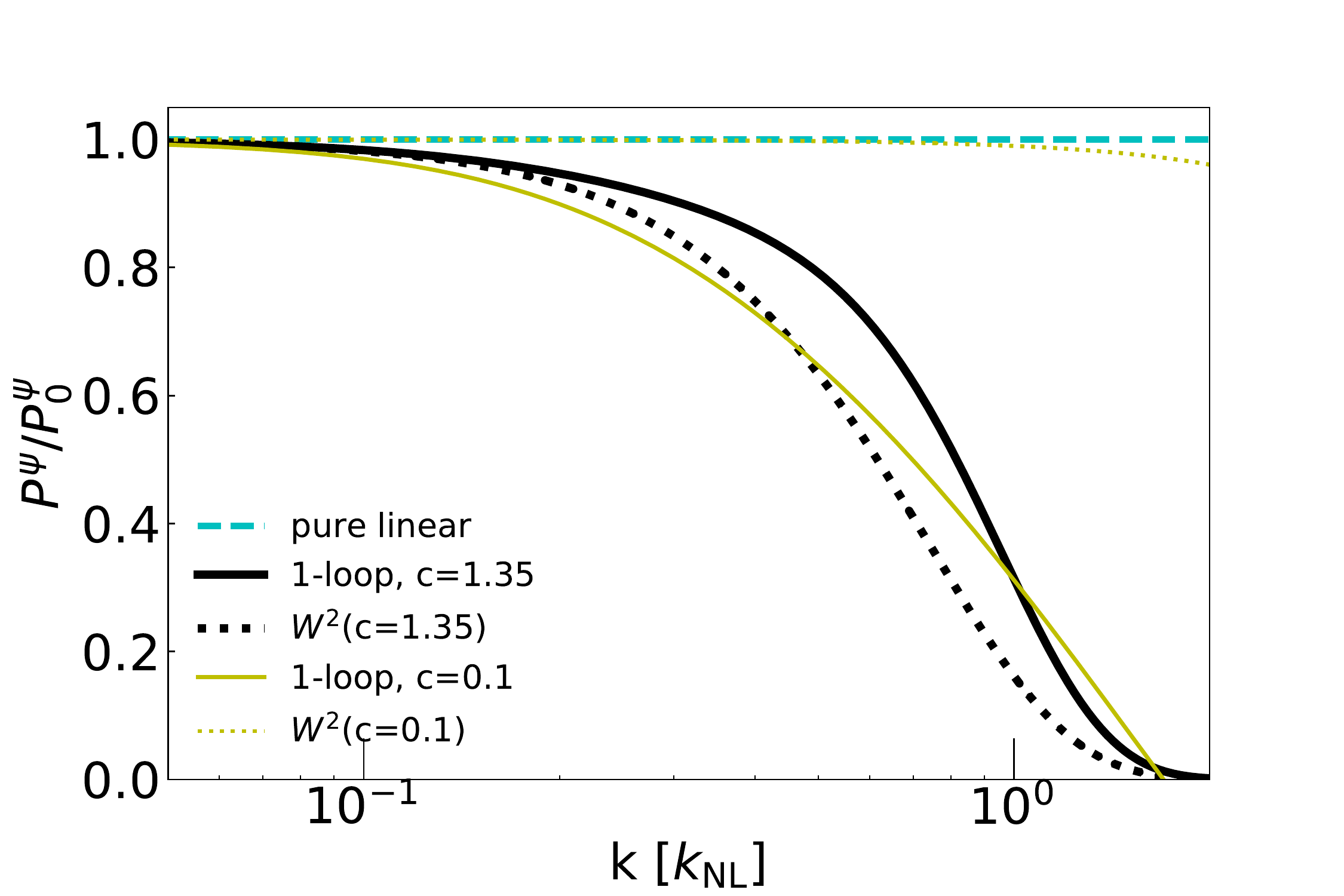}
 \end{center}
 \caption{
One loop suppression of displacement power, in 1D, for $n=0$.
}
 \label{fig:Ppsi0}
\end{figure}

For $W(k)=\exp\left(-k^2 c^2/2\right)$, the low-$k$ limit is 
$P^{(c)}_\psi(k\rightarrow 0,\eta) = c^2 k^2 P^g_\psi(k,\eta)$. This 
is a natural counterterm for the coefficient computed
by Eq. (\ref{eq:lowkdamp}), i.e., we can set 
$(P^{(a)}_{1,\psi} + P^{(c)}_{1,\psi} )(k\rightarrow 0,\eta=0) = 0$ by 
choosing
\be
c^2 = 
\frac{3}{2}\int d\eta^\prime
\left[\vL_0^{-1}\right]_{\psi\vupsilon}
\left(0,\eta^\prime\right)
e^{\eta^\prime}
\int dq^\prime
q^{\prime 2}
\frac{e^{-\frachalf \frac{ q^{\prime 2}}{\sigma^2(q^\prime,\eta^\prime)}}}
{\sqrt{2 \pi \sigma^2(q^\prime,\eta^\prime)}} ~.
\label{eq:csqcalibration}
\ee 
This matching is self-regulating, i.e., it is an
equation of the form $c^2=f(c)$, because 
$\sigma^2(q^\prime,\eta^\prime)$ depends on $c$. 
The formal divergence of the un-damped
$\sigma^2(q)$ for $n\geq 1$ is not a problem---it should be seen as an 
artifact of the unphysical nature of pure Zel'dovich displacements. 
The solutions are $c=$(0.6, 0.62, 0.74, 1.35)$\kNL^{-1}$ for 
$n=$(2, 1, 0.5, 0), respectively. (This matching is similar to but appears to
differ
somewhat from that in \cite{2015JCAP...11..049B}, in that they relied on the
fact that calculation of the term to match converges
for a linear theory $\Lambda$CDM power spectrum, 
rather than using the damping in their theory to cut it off.)
Figures \ref{fig:Ppsi2}-\ref{fig:Ppsi0} show the final 
displacement power spectrum, 
relative to pure linear.     
A key thing to notice in these figures 
is the dependence of the power suppression on
$c^2$. 
The larger value of $c^2$ results in {\it less} total suppression at 1-loop
order, even though the direct leading order effect is more suppression. At 
small $c^2$, there is {\it too much}    
stream crossing at leading order, which 
leads to large 1-loop power suppression. The larger $c^2$ introduces an 
appropriate level of suppression at leading order, allowing 1-loop 
corrections to be small. Using an even larger value of $c^2$ than we plot here
reverses the trend, giving leading order suppression too large so that the
1-loop correction must become large and positive to compensate.

This use of $W(k)$ with $c^2$ calibration is not the main point of the paper
(in case it is not clear: it is not fundamentally related to the 
inclusion of stream crossing)
so we defer most discussion to Sec. 
\ref{sec:discussion}, but we address a couple
questions here. First note that, at least in this 1D calculation, $c^2$ is
guaranteed to be positive, so we do not need to worry about $e^{-c^2 k^2}$ 
diverging. As discussed below, we expect any serious 3D application
to use an at least slightly more sophisticated replacement for the procedure 
used here, so there 
is no point in thinking about when the exact approach here might 
pathologically break down.
Second, it would be very natural for $c^2$ to be
time dependent, i.e., we would evaluate Eq. (\ref{eq:csqcalibration}) at
all $\eta$ instead of $\eta=0$. For an arbitrary power spectrum this would 
introduce significant complication as we would need to solve the matching for a
continuous function, with values at a given time depending on all values in 
the past, but 
for power law EdS models we would still only need to solve for one 
number because we know 
exactly
what the time dependence must be. The only scale in the problem 
is the nonlinear scale
where $\Delta^2(\kNL)=1\propto e^{2 \eta} k^{n+d}$, so 
$\kNL(\eta)\propto e^{-2\eta /(n+d) }$ (assuming $n+d>0$). $c^2(\eta)$ must be 
$\propto \kNL^{-2}(\eta)\propto e^{4\eta /(n+d)}$, 
with only the coefficient to be determined by matching. 
We expect that using this would make calculations at a given order more accurate
by bringing the leading order closer to the truth, but it is important to 
understand that the calculation is mathematically consistent either way, 
because in either case the part that is taken away is restored as a
perturbation. If we wanted to use a constant $c^2$ to make predictions in a
model with arbitrary power spectrum, there would be nothing fundamentally 
wrong with simply computing a different value at each redshift we were 
interested in---in each case the fixed $c^2$ used in the calculation would
represent a reasonable $\sim$time-averaged value for that redshift. There is 
certainly ambiguity here, but that is true of all perturbation theory,
where the definition of orders is generally not unique---what
we hope for is that the series we are computing converges to the
same answer for any reasonable choice.   

\subsubsection{Density power spectrum}

The nontrivial density power spectrum contribution is
\begin{eqnarray}
\left<\delta(\vk_1,\eta)\delta(\vk_2,\eta)\right>^{(a)}_1 &=& 
\int \dq_1 
\int \dq_2 ~e^{i \vk_1\cdot \vq_1+i \vk_2\cdot \vq_2}
e^{i \vk_1 \cdot\partial_{\vj_\vpsi(\vq_1,\eta)}+
i \vk_2\cdot \partial_{\vj_\vpsi(\vq_2,\eta)}}
Z_1^{(a)}(\vj) \left. \right|_{\vj=0} 
\\ \nonumber 
&=& \frac{3}{2} (2 \pi)^d\delta^D(\vk_1+\vk_2)
 \int \dq_{12} \dq d\eta^\prime \int\dq^\prime \dk 
\frac{\vk\cdot \vL^{-1}_{\psi\upsilon}\left(\vq\right)\cdot \vk_1}{k^2}
e^{i \vk_1\cdot \vq_{12}} \\ \nonumber 
& & \qquad \left[
e^{i \vk\cdot\vQ\left(\vq,\vq^\prime,\eta^\prime\vq_{12},\eta,\vk_1\right)}
-
e^{-i \vk\cdot\vQ\left(\vq,\vq^\prime,\eta^\prime\vq_{12},\eta,\vk_1\right)}
\right]
e^{-\frachalf k_i\sigma^2_{ij}(\vq^\prime,\eta^\prime)k_j } 
e^{-\frachalf k_{1 i} \sigma^2_{i j}(\vq_{12},\eta)k_{1 j}}
- {\rm mean~part}
\end{eqnarray}
where
$\vQ\left(\vq,\vq^\prime,\eta^\prime\vq_{12},\eta,\vk_1\right)
\equiv \vq^\prime-i \vXi\left(\vq,\vq^\prime,\eta^\prime,\vq_{12},\eta\right)
\cdot\vk_1$ 
and
\begin{eqnarray}
\vXi(\vq,\vq^\prime,\eta^\prime,\vq_{12},\eta)&\equiv&
\vC_\psi(\vq,\eta^\prime,\eta) -
\vC_\psi(\vq+\vq_{12},\eta^\prime,\eta)
-\vC_\psi(\vq+\vq^\prime,\eta^\prime,\eta)+ 
\vC_\psi(\vq+\vq^\prime+\vq_{12},\eta^\prime,\eta) ~.
\end{eqnarray}

In 1D we can again integrate over $k$:
\begin{eqnarray}
\left<\delta(k_1,\eta)\delta(k_2,\eta)\right>^{(a),1D}_1 
&=& 
-(2 \pi)\delta^D(k_1+k_2) \frac{3}{2}{\rm Im} \int dq_{12} dq d\eta^\prime 
e^{i k_1 q_{12}}
\vL^{-1}_{\psi\upsilon}(q,\eta,\eta^\prime) \\ \nonumber
& & \qquad \qquad
k_1\int dq^\prime{\rm erf}\left[\frac{q^\prime-i k_1
\Xi(q,q^\prime,\eta^\prime,q_{12},\eta)
}{\sqrt{2}\sigma\left(q^\prime,\eta^\prime\right)}\right] 
e^{-\frachalf k_1^2 \sigma^2(q_{12},\eta)} - {\rm mean~part}
\label{eq:Pd1Da}
\end{eqnarray}

The suppressed-power-restoring contribution to the  density power spectrum is
\begin{eqnarray}
\left<\delta(\vk_1,\eta)\delta(\vk_2,\eta)\right>^{(c)}_1 &=&
\int \dq_1
\int \dq_2 ~e^{i \vk_1 \cdot \vq_1+i \vk_2 \cdot \vq_2}
e^{i \vk_1 \partial_{\vj_\vpsi(\vq_1,\eta)}+
i \vk_2 \partial_{\vj_\vpsi(\vq_2,\eta)}}\left[
Z_1^{(c)}(\vj) -Z_1^{(c)}(0) Z_0(\vj)\right] \left. \right|_{\vj=0}
\\ \nonumber
&=&
\int \dq_1
\int \dq_2 ~e^{i \vk_1\cdot \vq_1+i \vk_2 \cdot \vq_2}
e^{i \vk_1 \partial_{\vj_\vpsi(\vq_1,\eta)}+
i \vk_2 \partial_{\vj_\vpsi(\vq_2,\eta)}}
 \vj^t \left(1-W\left(k\right)\right) \vC \vj~ Z_0(\vj)
\left. \right|_{\vj=0}
\\ \nonumber
&=&
-(2 \pi) \delta^D(k_1+k_2)
\int \dq
~\cos\left(\vk_1\cdot \vq\right) ~k^m_1\sigma_{1-W}^{2 mn}(q,\eta)
k^n_1 ~e^{-\frachalf k^i_1 \sigma_{ij}^2(q,\eta)k^j_1}~.
\end{eqnarray}

\begin{figure}[ht!]
 \begin{center}
  \includegraphics[scale=0.8]{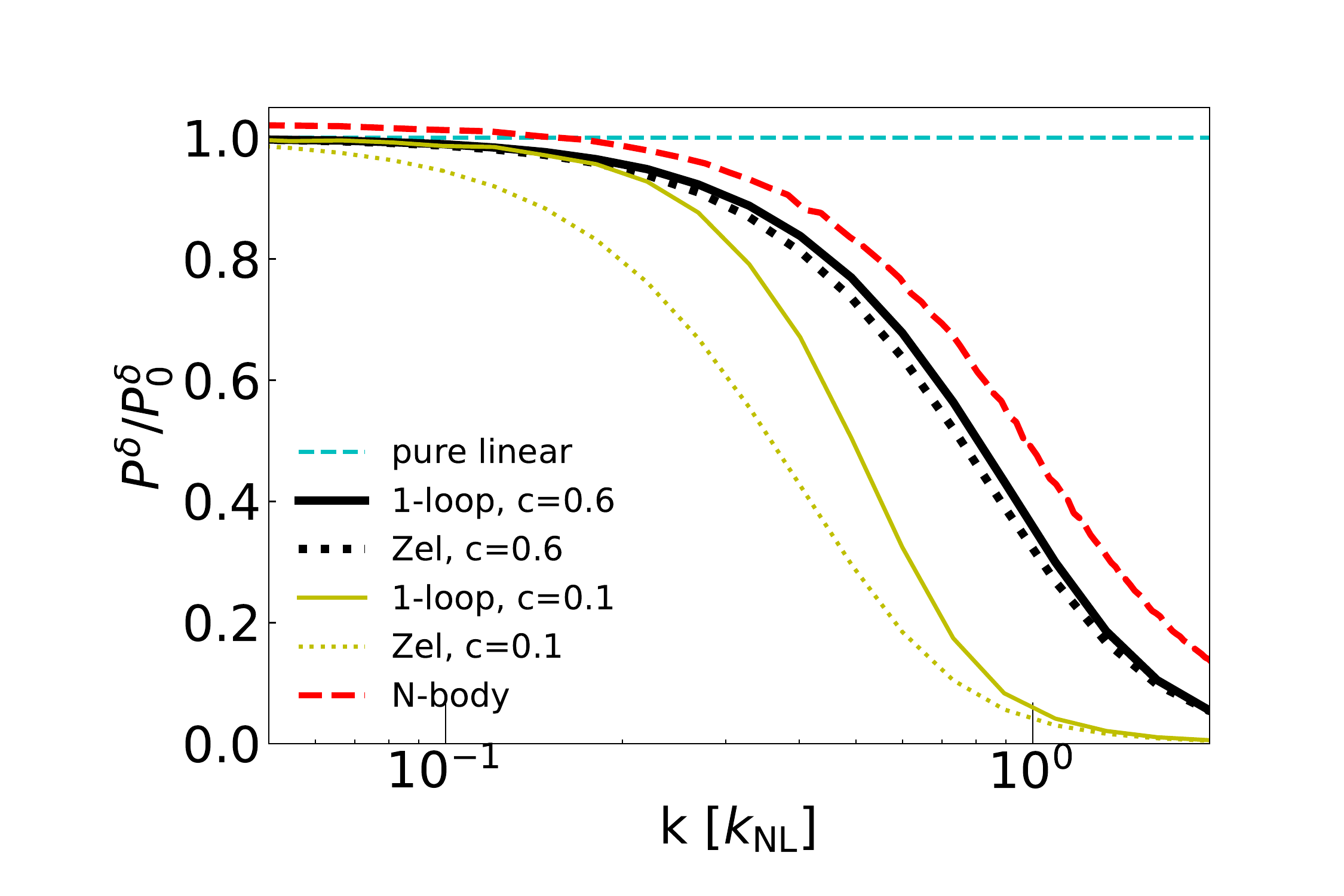}
 \end{center}
 \caption{
One loop density power, in 1D, for $n=2$, relative to the initial power law.
Solid black shows the final one loop prediction, using the value of $c$ 
fixed internally by the displacement power calculation as shown in 
Fig. \ref{fig:Ppsi2}.
Black dotted shows the Zel'dovich power for the same $c$, i.e., the difference
is the happily small one loop correction.  
Red dashed are the N-body results of \cite{2016JCAP...01..043M},
who say that their $n=2$ simulations had not fully 
converged. E.g., the points at low $k$ are almost surely a couple percent 
lower where they would agree well with our prediction (see the 
smaller-$n$ figures that follow where they did not have this issue). 
For comparison, we show the same quantities for $c=0.1$, where the one loop
calculation makes a valiant effort to correct the 
massive failure of Zel'dovich, but
it is clear that $c=0.6$ is a much better starting point for a perturbative
expansion. 
}
 \label{fig:Pd2}
\end{figure}

\begin{figure}[ht!]
 \begin{center}
  \includegraphics[scale=0.8]{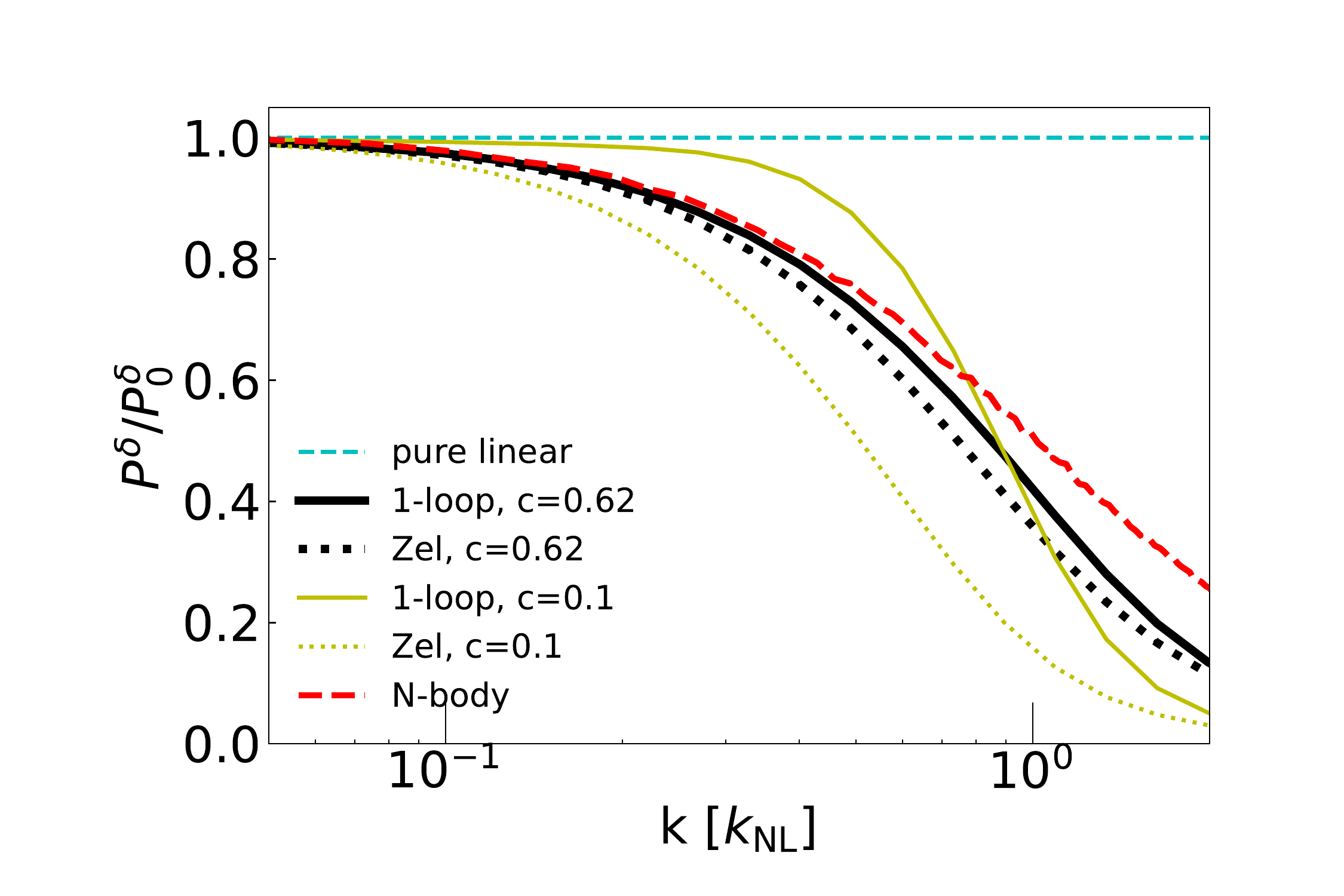}
 \end{center}
 \caption{
One loop density power, in 1D, for $n=1$, relative to the initial power law,
similar to Fig. \ref{fig:Pd2}.
Note that the values of $c$ used in all figures were computed
literally ``blind'' to the simulation 
results---it was clear that $W(k)$ was needed to keep blue
spectra from diverging (and from common-sense understanding of the physics), 
and the exact numerical values used here were settled
by the displacement power calculation 
before the density power was ever computed.
}
 \label{fig:Pd1}
\end{figure}

\begin{figure}[ht!]
 \begin{center}
  \includegraphics[scale=0.8]{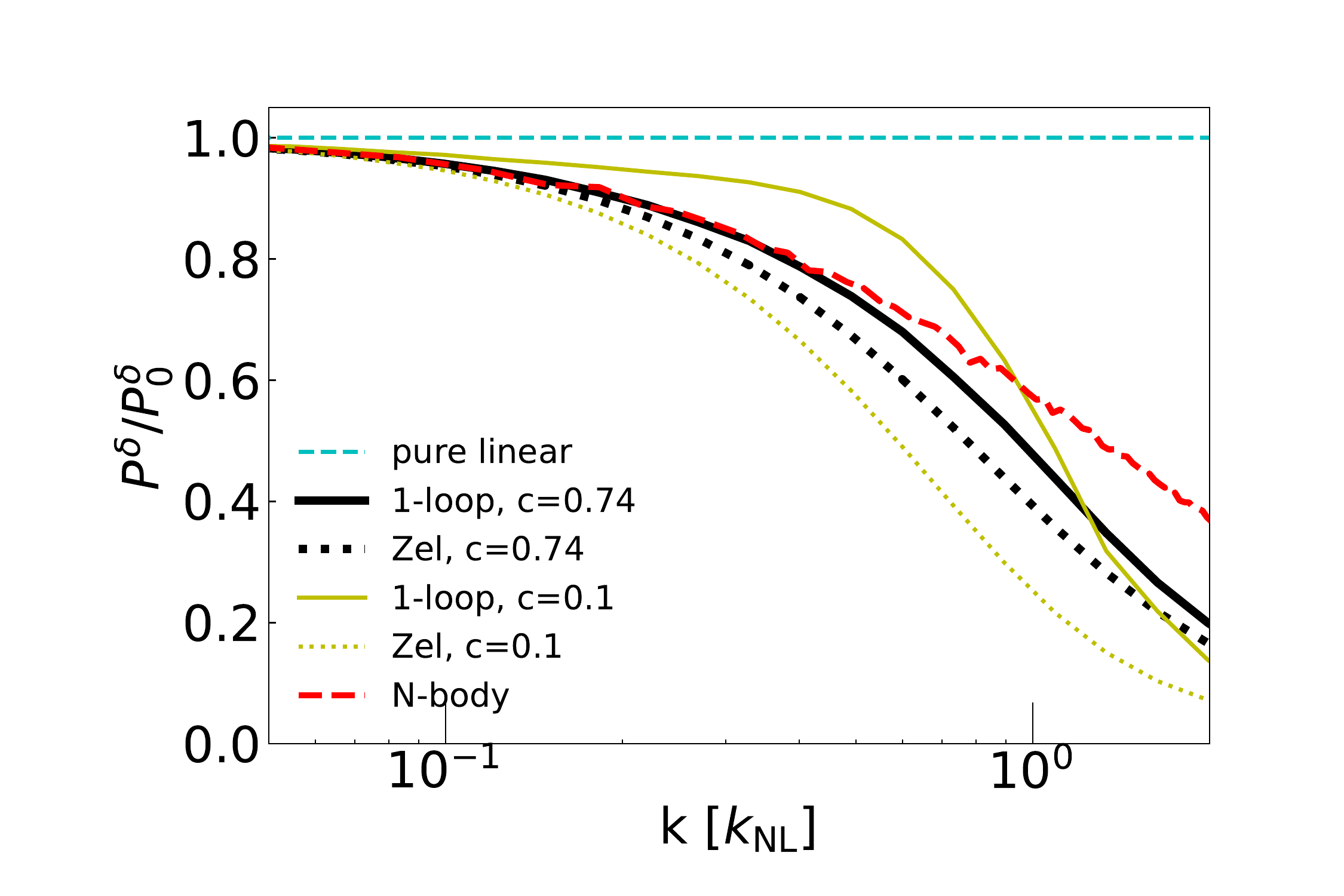}
 \end{center}
 \caption{
One loop density power, in 1D, for $n=0.5$, relative to the initial power law,
similar to Fig. \ref{fig:Pd2}. 
}
 \label{fig:Pd0.5}
\end{figure}

\begin{figure}[ht!]
 \begin{center}
  \includegraphics[scale=0.8]{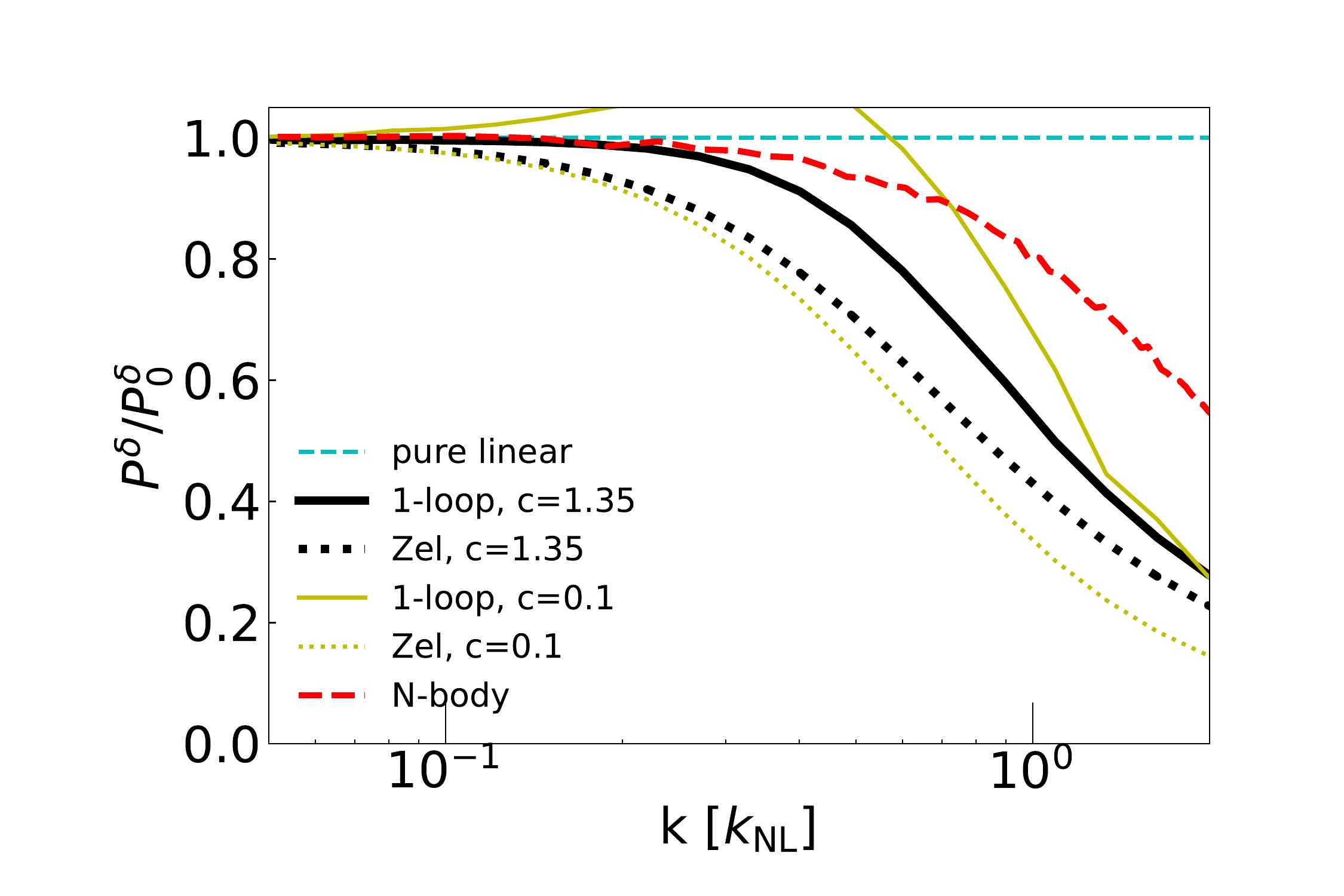}
 \end{center}
 \caption{
One loop density power, in 1D, for $n=0$, relative to the initial power law,
similar to Fig. \ref{fig:Pd2}.
}
 \label{fig:Pd0}
\end{figure}

So our final density power spectrum predictions are shown 
in Figs. \ref{fig:Pd2}-\ref{fig:Pd0}.
We see first that the numerical values of $c$ computed by setting the
one loop correction to the low-$k$ displacement power to zero do a great job
producing a perturbative expansion where the corrections are in fact small.
This calibration is entirely internal to the theory. Even if we only had the
Zel'dovich vs. corrected curves for density power here, it would be obvious 
that
the calibrated values of $c$ are a lot better than significantly different 
ones, just based on the principle that higher order terms in a
perturbative expansion should be small. 
Compared to the N-body results of \cite{2016JCAP...01..043M}, 
we see that the
predictions for $n=1$ and $n=1/2$ are excellent at low $k$, breaking down at 
the 10\% level at slightly higher $k$
than the free-parameter-fitted ``EFT'' curves shown in 
\cite{2016JCAP...01..043M}, and breaking down much more gracefully (in all
cases gradually under-predicting the simulation power).
The comparison for $n=2$ is murky because \cite{2016JCAP...01..043M} had
trouble achieving convergence in the simulations---in any case, considering
that the naive Zel'dovich prediction is divergent for $n=2$, our theory
seems like a very big step forward.  
Finally, $n=0$ presents a somewhat greater challenge for the theory. 
While we successfully predict that the power will only start deviating from
linear at higher $k$ than Zel'dovich, we do not predict these deviations 
very well once they happen. It may be that this is because a Gaussian $W(k)$
is less appropriate for this spectrum with more low-$k$ power. We see in
Fig. \ref{fig:Ppsi0} that our kernel matched in the asymptotic low-$k$ limit is 
not doing a good job matching the predicted damping at higher $k$. This 
can be fixed by finding a kernel that matches the calculated suppression 
better. There are many other possible refinements of the leading order theory
that might improve the results, discussed in Sec. \ref{sec:discussion}, but 
of course we may simply need another loop. 

\section{Discussion \label{sec:discussion}}

There are many obvious next steps for this formalism, e.g., computing 
displacement, density, etc. power spectra in 3D and comparing to 
numerical simulations \cite{2009PhRvD..80d3531C,2012JCAP...04..013T,
2007PASJ...59.1049N,2008JCAP...10..031N,
2008MNRAS.389.1675T,2009ApJ...693.1404S,
2009PASJ...61..321N,2011PhRvD..84f3501O}, 
computing other statistics like the bispectrum  
\cite{2007PhRvD..76h3004S,2007MNRAS.382.1460P,2012JCAP...06..021R,
2012PhRvD..86h3540B,
2012JCAP...06..018R,2016PhRvD..93h3517L,
2017arXiv170808941H}, or 
power spectrum covariance matrix 
\cite{2008MNRAS.384.1221N,2009ApJ...700..479T,
2011ApJ...726....7T,2016PhRvD..93l3505B}, etc.
Efforts to speed up PT numerical evaluation
\cite{2016PhRvD..93j3528S,2016PhRvD..94j3530S,2016JCAP...09..015M,
2017JCAP...02..030F,2017arXiv170808130S}
could be reconsidered in this context.
Here we discuss a few possibilities where we may have something useful to 
say about the path forward. 

\subsection{Non-Gaussianity, modified gravity, neutrinos, 
baryons, bias, redshift space distortions, 
etc.}

It should not be too hard to fit primordial non-Gaussianity 
\cite{2008PhRvD..78l3519M,2010JCAP...03..011B,
2010PhRvD..81f3530G,2011MNRAS.417L..79G,2014arXiv1412.4671A,
2017arXiv170806473M} into this 
formalism. By definition the modification of the starting point statistical 
distribution for $\vep$, Eq. (\ref{eq:epsilonlikelihood}), will add an extra
non-Gaussian part to the perturbative piece $S_p$. Note that any kind of 
polynomial in $\phi \propto k^{-2}\delta$ can be constructed by repeated 
applications of the $\delta$-generating operator.    

Typical modifications of 
gravity \cite{2014PhRvD..90l3515T,2017arXiv170510719A} 
should not present any new problem for this formalism.  
In standard perturbation theory, calculations are usually
simplified by the good approximation for density statistics 
\cite{2008PThPh.120..549T} 
that the growth at each order $n$ is proportional to $D^n(t)$,
where $D(t)$ is the linear growth factor. 
This means, as far as density statistics is considered, we have not needed to do numerical 
time integrals as part of PT [other than to compute $D(t)$]. 
For momentum correlators this approximation starts to be less than percent
accurate and corrections need to be taken into account  \cite{2016PhRvD..94f3516F}.
In the calculations of this 
paper, however, we have abandoned this feature---time integration is an
unavoidable part of the calculation including stream crossing. We also 
integrate over $k$-dependence of the propagator, which is not necessary in 
SPT. Having accepted these necessities, we can implement a modified 
gravity model that forces them on us with no additional complication.
Non-linear effects can be included as perturbations. 

Perturbation theory for massive neutrinos
\cite{2009JCAP...06..017L,2009PhRvD..80h3528S,
2015JCAP...03..046F,2010PhRvD..81l3516S,2010PhRvD..82h9901S}
may be an ideal application for this 
formalism.  The only practical difference at late times 
between neutrinos and cold dark matter 
amounts to initial conditions---the initial conditions for neutrinos include
large velocities that are uncorrelated down to arbitrarily small scales. 
It should be possible to represent the neutrinos with a displacement field
just like we have discussed for CDM by imagining a snapshot at some early
time and defining $\vN$ to give appropriate random velocities in addition to
the usual initial correlated perturbations. The first calculation to do
will be the displacement power spectrum to determine the small-scale
suppression kernel $W(k)$ for the neutrinos, which
will be appropriately longer range.
The key point is: because the formalism already fully includes stream 
crossing, and deals gracefully with naively very large dispersion in the 
initial conditions (e.g., infinite for $n=2$ here), there should be no need 
for, e.g., special evolution equations for neutrinos. To be clear: we would
now have separate $\vpsi$ and $\vupsilon$ for the neutrinos, driven by the
combined density field, with $\vL$ becoming a $4\times 4$ matrix multiplying
fields, etc..   
To more faithfully represent all the relevant physics,
this scenario may provide extra motivation for more detailed renormalization 
of $\vL$ and
renormalization of $\vN$, as discussed below.   

Unfortunately, what we have here is a theory for dark matter, which is 
not generally observable. Gravitational 
lensing \cite{2017arXiv170801538T,2017arXiv170801530D,
2016MNRAS.459.4467B,2014JCAP...09..024J,
2017arXiv170706948Z,2017JCAP...03..016D,2015JCAP...01..022M} 
is typically cited as the most 
direct tracer, but even this is sensitive to baryon pressure effects on 
small-scales \cite{2015MNRAS.454.2451E}, and ultimately is an observation of
galaxies (or some other source of photons), not directly mass, with 
corresponding biasing issues like intrinsic alignments 
\cite{2013JCAP...12..029C,2016MNRAS.456..207K,2017arXiv170809247B}.
Baryon displacement fields can easily be introduced to at least model the 
effect of the difference in initial conditions on large scales
\cite{1995ApJ...455....7M,1995ApJ...444..489H,1996ApJ...469..437S,
2002ApJ...569....1S,2003MNRAS.344..481Y,
2010PhRvD..82h3520T,2013PhRvD..87d3530B}.  
We could easily add explicit 
pressure terms for a given equation of state (this might 
be especially interesting for describing the \lyaf\
\cite{2000ApJ...543....1M,2002ApJ...580...42M,
2003ApJ...585...34M,2003ApJ...594L..71A,2003MNRAS.344..776M,
2005ApJ...635..761M,2006ApJS..163...80M,2007PhRvD..76f3009M}, 
where little effort has
been made to use higher order perturbation theory, even though its 
generally weak non-linearity would seem to make it a good candidate), but it
is not clear that the perturbation theory as it stands will represent the
physics fully correctly, 
because baryon streams should really never cross (they should shock instead), 
while the perturbation theory might effectively allow them to sometimes, at
which point, e.g., the pressure force might be effectively pointing in the 
wrong 
direction. $W(k)$ the way we use it probably cannot entirely control
this problem. It might be necessary to introduce something that more 
completely stops the possibility of crossing (e.g.,
\cite{1989MNRAS.236..385G,
1988Natur.334..129K,1990MNRAS.247..260W,2011A&A...526A..67V}), although we
would want to be sure that this was a controlled modification of the 
perturbative expansion like our $W(k)$, not an arbitrary hack. Another 
possibility might be to use Eulerian-type fields for baryons while still using
displacements for dark matter, as in Eulerian hydrodynamic simulations
\cite{1992ApJS...78..341C,2014ApJS..211...19B}. While some kind of effective
model with free or externally fixed parameters is inevitably needed for
modeling temperatures any time they are affected by star formation or other
complex physics, it would be interesting to see if any progress could be made
including an explicit temperature field computed from a few simple principles 
to describe a relatively simple system like the IGM \cite{2001ApJ...562...52M}.

One of the most hopeful uses for perturbation theory is to organize our 
understanding of biasing models describing how galaxies 
and other observables trace dark matter 
\cite{2009JCAP...08..020M,2014MNRAS.444.1400N,2016arXiv161109787D,
2017arXiv170602362H}.
The usual Eulerian bias prescriptions \cite{2006PhRvD..74j3512M,
2007PhRvD..75f3512S,2009ApJ...691..569J,2009JCAP...08..020M,
2012PhRvD..86j3519C,2014PhRvD..90l3522S,2017MNRAS.464.3409B,
2017MNRAS.467.3993A}
should be straightforward to implement in this formalism, with correlations
involving powers of $\delta(\vx)$ and its derivatives computed by repeated
applications of the $\delta$-generating operator. 
For Lagrangian bias 
\cite{2008PhRvD..78h3519M,2011PhRvD..83h3518M,2011MNRAS.416.1703E}
we can always write some function of $\vphi(\vq)$ as a weight next to the 
delta function in Eq. (\ref{eq:deltax}), and the tracers will 
stream-cross just like dark matter. 

Redshift space distortions \cite{2008PhRvD..78h3519M,2009JCAP...10..007M,
2009JCAP...11..026M,
2009PhRvD..80l3503T,2010MNRAS.406..803B,
2010PhRvD..82f3522T,2011PhRvD..83h3518M,2011JCAP...11..039S,
2012JCAP...02..010O,2012JCAP...11..014O,
2012JCAP...11..009V,2012MNRAS.425.2099S,2012MNRAS.421.2656M,
2014JCAP...05..003O,2017PhRvD..95f3528C} 
can be included for dark matter with no additional 
approximation. 
We just add the appropriate apparent radial displacement 
$\propto \vupsilon_\parallel$ in the exponential when constructing density 
fields. Redshift space distortions for tracers like galaxies are a little 
trickier because, while low-$k$ modes of velocity should be the same as
dark matter, this is not generally
true on smaller scales, so some kind of generalized biasing model is
needed to allow for the full range of possibilities. 

\subsection{Deriving effective theories}

Suppose that instead of statistics we really want a theory for the Eulerian
field $\delta(\vx)$. We can introduce it into $Z$ using a delta function 
like this:
\be
Z = \int d\vdelta_S~ d\vphi ~d\vchi~
\delta^D[\vdelta_S-\vdelta_S(\vpsi)]~ 
e^{-S(\vphi,\vchi)}
= \int d\vdelta_S ~d\vpi ~d\vphi ~d\vchi~
e^{i \vpi^t\left[\vdelta_S-\vdelta_S(\vpsi)\right] -S(\vphi,\vchi)}~,
\ee 
where we have written $\vdelta_S$ to indicate that we probably want to construct
a theory for a smoothed version of $\vdelta$.
We can now substitute Eq. (\ref{eq:deltak}) for $\vdelta(\vpsi)$ and if we can
perform the integrals over $\vphi$ and $\vchi$ we will have a theory for
$\vdelta_S$, with $\vpi$ playing the role that $\vchi$ does for $\vphi$.
We can perform these integrals perturbatively by pulling the standard linear
approximation out of $\vdelta(\vpsi)$ to write 
$\delta(\vk)=i \vk \cdot \vpsi(\vk) +...$ where the leading piece here will
become part of the Gaussian integral over $\psi$ and the rest will be 
included in the perturbative part of the integration. I.e., we can achieve a 
generating function for $\vdelta$ statistics that can be made increasingly 
accurate by including higher order perturbative integral contributions. 
The formula will be for $\delta(t)$, and so imply evolution 
equations (and initial conditions) which could be used in different ways.
If the calculation is done consistently for smoothed $\delta_S$, the evolution
equations will correctly represent dynamics entirely in terms of the smoothed 
fields. 
A similar calculation could be used to construct equations for other fields,
e.g., moments of the velocity distribution function (momentum, energy, etc.
\cite{2011JCAP...04..032M})
can easily be written as integrals like the one for $\vdelta(\vpsi)$.
One might ask ``how likely is it that this could be an accurate theory, 
when it includes integration over small-scale $\vphi$?'' That is not clear, but
at least the size of perturbative terms will allow an internal estimate, and
the $W(k)$-type suppression of small scales at leading order should aid 
convergence.

Another possibility, especially relevant if calculations in the full 
stream-crossing theory turn out to be slower than in traditional
PT, would be to calibrate the free parameters of previous theories 
(e.g., \cite{2016JCAP...12..007V}) 
by matching predictions to the stream crossing theory in the low-$k$ limit. 
It should be possible to make the matching step fast because formulas simplify
in the low-$k$ limit.  

\subsection{More sophisticated renormalization}

We intentionally tried to minimize the footprint of $W(k)$ in this paper, to 
avoid distracting from the main point about stream crossing, but clearly the
issue it is addressing (of traditional linear theory 
being a terrible starting point for
perturbation theory) is very generally important.  
Our approach here evolved from a more standard field theory starting point
\cite{1995iqft.book.....P}.
One of our original motivations was the observation that 
``EFT of LSS'' proponents 
\cite{2012JCAP...07..051B,2012JHEP...09..082C,2014PhRvD..89d3521H} did not seem to be taking their effective theory 
seriously 
enough, in that their prescription introduced terms intended to represent
effective pressure/viscosity, but treated them as perturbations to the
traditional fluid equations, rather than taking them seriously as part of the
leading order linear evolution as one naturally would have done if given a
system with these terms already in the linearized equations. This was
presumably for ease of computation, but means that the terms can only be used
to cancel potentially divergent parts of the higher order corrections, rather
than prevent the divergence from ever happening as they probably physically
should do. We originally thought to improve their treatment by fully including
$k^2 \delta$-type terms in $\vL$ where they would naturally damp the 
propagator 
\cite{2015JCAP...11..049B,2015JCAP...01..014R,2016JCAP...02..032F}.
We thought to follow a standard Wilsonian 
RG procedure \cite{1974PhR....12...75W,1995iqft.book.....P} of 
integrating out fields in shells of $\vk$, absorbing the corrections into 
the coefficients of the effective theory. E.g., computing 
$\left< \vchi \vphi^t\right>$ perturbatively will give you corrections to 
$\vL^{-1}$, which can then be disentangled from corrections to $\vN$ in 
$\left<\vphi\vphi^t\right>$. Higher order interactions will also be generated.
Because of the self-regulating nature of the propagator damping, i.e., that
increasing a coefficient like $c^2$ generally decreases the calculated 
small-scale contribution to it, we expected to find $c^2$
values like we obtained here as some sort of fixed points of the RG evolution.  
At some point we realized that this full procedure might not be 
necessary---that we could capture the 
most relevant physics in the simpler way presented
here. (We tell this historical story because it is helpful to understand the
relation between our approach and standard field theory approaches.)
A natural thought at this point would be ``ok, what you have is a 
poor-person's version of the full Wilsonian treatment, so a next step must be
to more fully implement that.'' But it is not clear that this is 
correct---it may be that, 
{\it for our problem}, standard field theory renormalization 
should be seen as the poor-person's version of what we do here.  
Renormalization in quantum field theory 
takes the form it does because of the infinities that are
apparently unavoidable in the calculations, i.e., where we had an equation
like $c^2=f(c)$ to solve, they have $c^2=f(c)+\infty$. Obviously solving this
directly is no good, so they go on to observe that the derivative of 
$c^2$ with respect to some matching point (like our $k\rightarrow 0$) or
similarly the contribution from integrating out a small shell of $\vk$ in
the Wilsonian picture, is finite. This allows them to
calculate how $c^2$ changes with scale (where note that here we are using
$c^2$ abstractly to represent any parameter, like a particle mass), but the
initial condition for this running is still corrupted by the infinity and
must be treated as a free parameter. The possible exception to this problem
is if the parameter hits a fixed point of the RG evolution, so it acquires
a value independent of the initial conditions, but this generally does not
happen for all parameters. If we do not have any true deep-UV sensitivity in 
our system, 
i.e.,
do not have any divergence once enough correct physics is included for the
system to self-regulate, we can hope to avoid free parameters, solving for
parameters of the Gaussian model as we have done here for $c^2$ instead 
of only computing relative 
values between different scales. From the 
conventional point of view this is probably equivalent to saying all parameters
flow to a fixed point. If this sounds like wishful thinking, note that
we implicitly take for granted that it is possible to construct an effective 
theory of dark
matter clustering that has no free parameters, while being finite and 
completely insensitive to small-scale degrees of freedom. We
use this theory all the time: N-body simulations, 
where we generally take for
granted that larger scale structure converges as the numerical
resolution increases. 

There is another reason to prefer the approach we use to more literally
pursuing the Wilsonian picture of integrating out shells of modes: in those
calculations, generally the integration results in a contribution that cannot
be represented by a convenient, small number of terms in the Lagrangian
(equivalent to terms in our $S$), 
so one is forced to ``truncate the basis,'' i.e., ignore some of 
the
calculation that does not project onto the terms you want to track. This is a
potential source of error, not necessarily easy to control, which we avoid in 
our approach here where we do not throw anything away. Maybe the two approaches
can be combined somehow---if nothing else, the exercise of integrating out 
small 
scale modes should help to identify and understand the useful modifications 
of the expansion. 

Our modification of $\vL^{-1}$ by a single overall suppression function $W(k)$  
with semiarbitrarily chosen Gaussian form
was intended to absolutely minimally capture what are of course really more
complicated effects. While the details should be filled in by higher order
terms, this should be more efficient if the
leading order model can be improved.
The first extension one might make is to allow for 
separate $\vpsi\vpsi$, $\vpsi\vupsilon$, and $\vupsilon\vupsilon$ kernels.
While there is always a temptation to judge the value of 
this kind of detail based on whether it appears to improve agreement
with simulations, this is really not necessary---these differences are
an inevitable consequence of 
computing all the elements of $\left<\vphi \vphi^t\right>$ and 
$\left<\vphi \vchi^t\right>$, and can be calibrated entirely 
internally, i.e., if we compute something other than 
the $\left<\vpsi \vpsi^t\right>$ that we computed here, using $c^2$ from 
$\left<\vpsi \vpsi^t\right>$, 
we would find a less perfect match. By making $W(k)$ a matrix in 
$(\vpsi,\vupsilon)$, we can match all terms in $\left<\vphi \vphi^t\right>$
at once. More generally, we should also match
$\left<\vphi \vchi^t\right>$, which is more directly identified with 
$\vL^{-1}$, while $\left<\vphi \vphi^t\right>$ is also sensitive to, and 
therefore can be used to make corrections to, the noise $\vN$. The noise 
modifications should presumably be additive instead of multiplicative and
go like $k^4$ at low $k$ \cite{1980lssu.book.....P}. We could also attempt 
to match computed corrections to higher $k$ by modifying the form of 
$W(k)$ [and similarly $\delta N(k)$ if using it]. Similarly, these 
modifications are generally matrices in time, which can be calibrated in more
detail by evaluating unequal time correlations.
Also similarly, in higher dimensions $W(k)$ could be generalized to be 
asymmetric
depending on the relation between its vector indices and $\vq/\vk$
(like the dependence of $C_{\psi_i \psi_j}(\vq)$ on whether the indices 
are along or transverse to $\vq$). In other words:
ultimately $\vL^{-1}$
and $\vN$ are general matrices in the full relevant space, and one could 
imagine working to make all loop corrections to them zero. 
Note, however, that we
should not think that each new correction absorbed will make the final 
predictions more accurate by the full amount 
absorbed---once the corrections are small 
enough to take the form of a rapidly converging series, it should not matter
much if they are absorbed at lowest order or computed perturbatively. 
To identify non-Gaussian terms, it will probably be useful to perform a
large/small scale split of the fields, perturbatively integrate out the small 
scale part, and
see how to interpret the result as a modification of the large-scale equations
(this would also tell us how to modify the Gaussian part, if we could not 
guess). 

We tried to keep our $W(k)$ calibration simple by including the 
contribution to 
changing $P_\psi$ from all scales, but, 
especially if we were isolating $\vL^{-1}$ by computing
$\left<\vphi \vchi^t\right>$, we might encounter an
issue known in more standard perturbation theory, that very large scale bulk 
flows
can make significant (even divergent) contributions to matching
calculations like we did here even though their effect on smaller, 
observation-scale fluctuations
is not actually to damp them. 
In fact, for an asymptotically large-scale flow 
there should be no effect at all, by Galilean invariance. 
Even very large scale stream 
crossing will not necessarily have a simple damping effect on smaller
scale fluctuations within the flow (it seems there will be a change in 
background density, plus noise from the uncorrelated fluctuations in the other
stream, but not the direct damping you get from displacements streaming out 
of their own driving perturbations).
If this is found to be an issue, it will probably be useful in the future to 
do something like splitting the fields
into a small and large scale contribution, basically relative to the scale of
observation, and only including the contribution from the small-scale part when 
modifying $\vL^{-1}$. Because we are not permanently 
throwing anything away in these 
calculations, only shifting some of the naive Gaussian part of the functional
integral into the perturbative part, the results should not be very 
sensitive to, e.g., where exactly you put the large/small split. 

Finally, note that some attempts to implement Wilsonian RG 
\cite{2007JCAP...06..026M,2007PhRvD..76h3517I,2008MPLA...23...25M,
2017JCAP...01..048F}
ideas have been criticized for 
only integrating out small-scale fluctuations in the initial conditions,
not in the evolving fields \cite{2008JCAP...01..029R}. 
It is clear in the formalism presented
here how one could integrate these fluctuations out of all fields at all 
times.  

We thank Matt McQuinn for the N-body results from \cite{2016JCAP...01..043M},
and \anze\ Slosar and Zack Slepian for helpful comments on drafts. 

Z.V.~is supported in part by the U.S. Department of Energy contract to SLAC no.~DE-AC02-76SF00515.

\bibliography{cosmo,cosmo_preprints}

\appendix

\section{Pedagogical explanation of the key math trick \label{app:mathtrick}}

The very 
simple math trick that makes this paper work may not be easy to identify within
the formalism of our calculations, so we attempt to explain it more 
pedagogically here. Standard perturbation theory amounts to 
repeatedly performing the following type of integral:
\be
\int dx~ e^{-\frachalf x^2 +j x - \lambda x^3+...}=
\int dx~ e^{-\frachalf x^2 +j x}\left(1 - \lambda x^3+...\right)
\ee
i.e., it is based on the fact that one kind of integral we can do
analytically is a Gaussian times a polynomial. If some more general function
of $x$ appears in the exponential here,
one's first instinct is to Taylor expand it 
to obtain a polynomial. 
However, we observe that we can also do
\be
\int dx~ e^{-\frachalf x^2 +j x - \lambda e^{\alpha x}+...}=
\int dx~ e^{-\frachalf x^2 +j x}\left(1 - \lambda e^{\alpha x}+...\right)
=\int dx~ 
e^{-\frachalf x^2 +j x}
-\lambda e^{-\frachalf x^2 +(j+\alpha) x}
+...
\label{eq:mathtrick}
\ee
i.e., if the coefficients of the Gaussian include exponentials, we can simply
absorb them back into the Gaussian and again do the integral analytically. 
In the main calculation in this paper the formula for density as exponential in
displacement, Eq. (\ref{eq:deltak}), is analogous to $e^{\alpha x}$ here. We
avoid losing stream crossing by {\it not} Taylor 
expanding $e^{i \vk \cdot \vpsi}$
(which is tempting because it gives us a polynomial in $\vpsi$), 
but instead expanding this whole exponential factor out of the 
$e^{-S}$ exponential and doing the calculation 
in the way of Eq. (\ref{eq:mathtrick}). 
The ``action of the derivative operator'' that leads to Eq. (\ref{eq:Z1a}) 
amounts to, underneath all the notation, 
an integration over force derived from 
density written in terms
of exponentiated displacement. This is the key to the new result. 

\section{Jacobian of $\vep \rightarrow \vphi$ variable change 
\label{app:Jacobian}} 

Our evaluation of the Jacobian determinant factor 
in the change of variables from $\vep$ to $\phi$ follows similar 
calculations done in \cite{2001A&A...379....8V,2007A&A...465..725V}.
\be
{\rm det}\left( \frac{\partial \vep}{\partial \vphi}\right) =
{\rm det}\left(\vL_0 +\frac{\partial \vDelta_0(\vphi)}{\partial \vphi}\right) =
{\rm det}\left(\vL_0\right)
{\rm det}\left(\vI+
\vL_0^{-1}\frac{\partial \vDelta_0(\vphi)}{\partial \vphi}\right) 
\propto e^{{\rm Tr}\ln \left(\vI+\vL_0^{-1}\vM\right)}
\ee
where we have dropped the obviously field-independent 
${\rm det}\left(\vL_0\right)$, defined
$\vM \equiv \frac{\partial \vDelta_0}{\partial \vphi}$, and used the 
identity ${\rm det}\vA = e^{{\rm Tr}\ln \vA }$.
Note that $\vDelta_0$ has only an $\vupsilon$ component, but only depends on 
$\vpsi$, which means that in $\vM$ only 
$\vM_{\vupsilon \vpsi}$ is non-zero. 
Taylor expanding in $\vL_0^{-1}\vM$ gives
\be
{\rm Tr}\ln \left(\vI+\vL_0^{-1}\vM\right)
=-\sum_{n=1}^{\infty} \frac{1}{n}
{\rm Tr}\left[\left(-\vL_0^{-1}\vM\right)^n\right]~.
\ee
All of these terms take the form
${\rm Tr}\left[\vL^{-1}_{0 \vpsi\vupsilon} 
\vM_{\vupsilon\vpsi} \vL^{-1}_{0 \vpsi\vupsilon} 
\vM_{\vupsilon\vpsi} ...\right]$. With $\vM$ a delta function and
$\vL_0^{-1}$ a Heaviside function in the difference between their time 
indices, we see that the trace, which 
requires time to flow in a circle, i.e.,   
$\int dt_1 dt_2... dt_n 
A(t_1,t_2)...A(t_n,t_1)$,  
must give zero except possibly for the tricky limit where all times 
are the same so all $\Theta$ functions are evaluated at zero (with one integral
over time remaining). For a general evolution equation, we would need a more
subtle analysis of the equal time limit, but here it is simple because
$\vL^{-1}_{0 \vpsi\vupsilon}(\eta,\eta)=0$ independent of the $\Theta$ 
function, so every term is unambiguously zero. 

\subsection{Alternative derivation}

We learned a different way to derive our basic 
functional integral starting point
from \cite{2016NJPh...18d3020B}. 
Starting from the likelihood function
for $\vep$, Eq. 
(\ref{eq:epsilonlikelihood}), we can enforce the evolution Eq.
(\ref{eq:compactev}) using a delta function:
\be
Z(\vj) = 
\int~ d\vphi ~d\vep ~\delta^D\left[\vphi-\vphi\left(\vep\right)\right]~
e^{-\frac{1}{2} \vep^t \vN^{-1} \vep}
\propto
\int~ d\vphi~ d\vep ~\delta^D\left[\vep-\vep\left(\vphi\right)\right] ~
e^{-\frac{1}{2} \vep^t \vN^{-1} \vep}
=
\int d\vphi ~d\vep~ d\vchi ~
e^{i\vchi^t \left[\vep-\vep\left(\vphi\right) \right]
-\frac{1}{2} \vep^t \vN^{-1} \vep}
\ee
where we have used the same fact that the Jacobian of the $\vep$-$\vphi$ 
transformation is field independent to change arguments in the delta 
function.  
Note that this transformation is invertible, e.g., if we imagine 
discretizing the 
evolution equation, it is easy to see that knowing $\vep$ uniquely 
determines $\vphi$ and vice versa (this is easier than inverting a 
final field to determine initial conditions---here we know the field at all
times by definition). We can now insert Eq. (\ref{eq:compactev}) 
for $\vep(\vphi)$ and
perform the Gaussian integral over $\vep$ to obtain 
Eq. (\ref{eq:addchi}-\ref{eq:Sphichi}). $\chi$ plays a role enforcing the
evolution equation instead of appearing through an apparently arbitrary 
Gaussian integral trick. 

\section{Low-$k$ expansion of density power formula}

We can consider a small $k$ expansion of the erf part
of Eq. (\ref{eq:Pd1Da}). For small $y$,
${\rm erf}(x+i y) = {\rm erf}(x)+2 i y \exp\left(-x^2\right)/\sqrt{\pi}+...$.
The leading term gives zero in the $q^\prime$ integral, so we have:
\begin{eqnarray}
- \frac{3}{2}{\rm Im} \int dq_{12} dq d\eta^\prime
e^{i k_1 q_{12}}
\vL^{-1}_{\psi\upsilon}(q,\eta,\eta^\prime)
k_1\int dq^\prime{\rm erf}\left[\frac{q^\prime-i k_1
\Xi(q,q^\prime,\eta^\prime,q_{12},\eta)
}{\sqrt{2}\sigma\left(q^\prime,\eta^\prime\right)}\right]
e^{-\frachalf k_1^2 \sigma^2(q_{12},\eta)}  \\ \nonumber
\simeq
3 k_1^2 \int dq_{12} dq d\eta^\prime
\cos\left(k_1 q_{12}\right)
\vL^{-1}_{\psi\upsilon}(q,\eta,\eta^\prime)
\int dq^\prime
\Xi(q,q^\prime,\eta^\prime,q_{12},\eta)
G\left[q^\prime,\sigma\left(q^\prime,\eta^\prime\right)\right]
e^{-\frachalf k_1^2 \sigma^2(q_{12},\eta)}
\\ \nonumber
\simeq
3 k_1^2 \int dq_{12}
\cos\left(k_1 q_{12}\right)
e^{-\frachalf k_1^2 \sigma^2(q_{12},\eta)}
\int
d\eta^\prime
[\vL^{-1}_0]_{\psi\upsilon}(\eta,\eta^\prime)
\int dq^\prime
\Xi_W(0,q^\prime,\eta^\prime,q_{12},\eta)
G\left[q^\prime,\sigma\left(q^\prime,\eta^\prime\right)\right]
\end{eqnarray}
where in the last line we note that the $q$ integral is a
convolution between $W(q)$ in $\vL^{-1}$ and $\Xi$ and remove the
$q$ integral by defining
$\Xi_W$ to be $\Xi$ with the power spectrum used to compute it multiplied
by an extra factor of $W(k)$. 

For the 1D cases we have tried, we find that,
while the
approximation is valid at low enough $k$, ``low enough'' is always too
low to be useful.

\end{document}